\documentclass[prb,twocolumn,aps,10pt]{revtex4-1}
\usepackage{amsmath}
\usepackage{amssymb}
\usepackage{enumerate}
\usepackage{amsthm}
\usepackage{graphicx}
\usepackage{hyperref}
\usepackage{comment}
\usepackage[caption=false]{subfig}
\newcommand{\ket}[1]{|#1\rangle}
\newtheorem{lemma}{Lemma}
\newtheorem{corollary}{Corollary}
\newtheorem{conjecture}{Conjecture}
\newcommand{\eqnref}[1]{Eq.\ (\ref{#1})}

\newcommand{\wSigma}{{\widetilde{\Sigma}}}
\newcommand{\wsigma}{{\widetilde{\sigma}}}

\begin{document}

\title{Crystalline topological phases as defect networks}

\author{Dominic V. Else}
\affiliation{Department of Physics, Massachusetts Institute of Technology,
Cambridge, MA 02139, USA}
\affiliation{Department of Physics, University of California, Santa Barbara, CA
93106, USA}

\author{Ryan Thorngren}
\affiliation{Department of Condensed Matter Physics, Weizmann Institute of Science, Rehovot, Israel}
\affiliation{Department of Mathematics, University of California, Berkeley, CA, 94720, USA}

\begin{abstract}
A crystalline topological phase is a topological phase with spatial symmetries. In this work, we give a very general physical picture of such phases: a topological phase with spatial symmetry $G$ (with internal symmetry $G_{\mathrm{int}} \leq G$) is described by a \emph{defect network}: a $G$-symmetric network of defects in a topological phase with internal symmetry $G_{\mathrm{int}}$. The defect network picture works both for symmetry-protected topological (SPT) and symmetry-enriched topological (SET) phases, in systems of either bosons or fermions. We derive this picture both by physical arguments, and by a mathematical derivation from the general framework of [Thorngren and Else, Phys.~Rev.~X \textbf{8}, 011040 (2018)]. In the case of crystalline SPT phases, the defect network picture reduces to a previously studied dimensional reduction picture, thus establishing the equivalence of this picture with the general framework of Thorngren and Else applied to crystalline SPTs.
\end{abstract}

\maketitle

A topological phase of matter is a gapped phase of matter that is characterized by patterns of quantum entanglement in the ground state, rather than by spontaneous symmetry breaking \cite{Wen1990}. A key aspect of topological phases is the interplay between the low-energy topological features and the microscopic symmetries of the system. In particular, systems which are smoothly connected in the absence of symmetry may become distinct phases when a symmetry is enforced, distinguished by the symmetry action on the topological degrees of freedom; these are called \emph{symmetry-protected topological} (SPT) \cite{Gu2009,Pollmann2010,Pollmann2012,Fidkowski2010,Chen2010,Chen2011,Schuch2011a,Fidkowski2011,CGLW,Chen2011b,Vishwanath2013,Wang2014, K,Gu2014,Else2014,Burnell2014,
Wang2015,Cheng2015} or \emph{symmetry-enriched topological} (SET) \cite{Maciejko2010,Essin2013,Lu2013,Mesaros2013,Hung2013,BBCW,Cheng2015a} phases, depending on whether in the absence of symmetry they are the trivial phase or not.

Of the realistic microscopic symmetries that can act on a quantum lattice model, we can divide them into two classes: an \emph{internal} symmetry (such as charge conservation, spin rotation or time-reversal)
acts locally on each site on the lattice, whereas a \emph{spatial} symmetry (such as spatial reflection or rotation) moves lattice sites around in space. We call a topological phase with spatial symmetry (or more generally, a symmetry group combining both internal and spatial symmetries) a \emph{crystalline topological phase} \cite{Fu2011,Hsieh2012,Dziawa2012,Tanaka2012,Xu2012,Zhang2013,
Fang2012,Fang2013,Chiu2013,Morimoto2013,Slager2013,Shiozaki2014,Chiu2016,
Isobe2015,Hsieh2014a,Qi2015,
Yoshida2015a,SS,Wen2002,Essin2013,
Hsieh2014,Cho2015,Yoshida2015,
BBCW,Lapa2016,You2014,Hermele2016,
Cheng2015a,Jiang2016}.
The development of the theory of crystalline topological phases, especially with strong interactions, has lagged behind the theory of topological phases with internal symmetry, despite their intrinsic interest. The reason is perhaps that, whereas purely internal symmetries can be understood simply in terms of an action of symmetry on the field theory describing the low-energy, long wavelength physics of the system, spatial symmetries relate to more microscopic properties of the underlying lattice.

Nevertheless, two competing general frameworks have emerged that are hypothesized  to give a general classification of crystalline topological phases. The first, stated in Ref.\onlinecite{SHFH} for point groups and then extended in Ref.\onlinecite{Huang_1705}, applies to \emph{invertible} topological phases. These are conjectured to be captured by layers of $k$-dimensional \emph{internal-symmetry} topological phases arranged in some spatial configuration in $d$-dimensional space, where $k$ ranges from $0$ to $d$. We call this the ``block state'' picture of crystalline topological phases.

The second framework was proposed by us in Ref.~\onlinecite{Thorngren_1612}. This framework works also for non-invertible phases, i.e.\ phases which contain non-trivial topological excitations such as anyons. We gave two physical pictures in Ref.~\onlinecite{Thorngren_1612}, which were argued to lead to the same classification. One picture was based on smooth states, which are states that vary very slowly in space, with a radius of spatial variation $R$ that is much larger than the lattice spacing and the correlation length (but nevertheless must be on the order of the unit cell size of the spatial translation symmetry, if it is present). The other picture was based on the idea that topological phases with symmetry should be distinguished by their responses to gauge fields, supplemented by a proposal for the meaning of gauging a spatial symmetry.  An important consequence of the framework of Ref.~\onlinecite{Thorngren_1612} is the \emph{Crystalline Equivalence Principle}, which states that the classification of phases with spatial symmetry $G$ is in one-to-one correspondence with the classification of phases with \emph{internal} symmetry $G$ (modulo some ``twists''; for example, a unitary but spatially orientation-reversing spatial symmetry such as reflection maps to an anti-unitary internal symmetry such as time-reversal). The same result was obtained from a tensor network point of view in the case of bosonic group cohomology SPTs in Ref.~\onlinecite{Jiang2016}.

\begin{figure}
\includegraphics{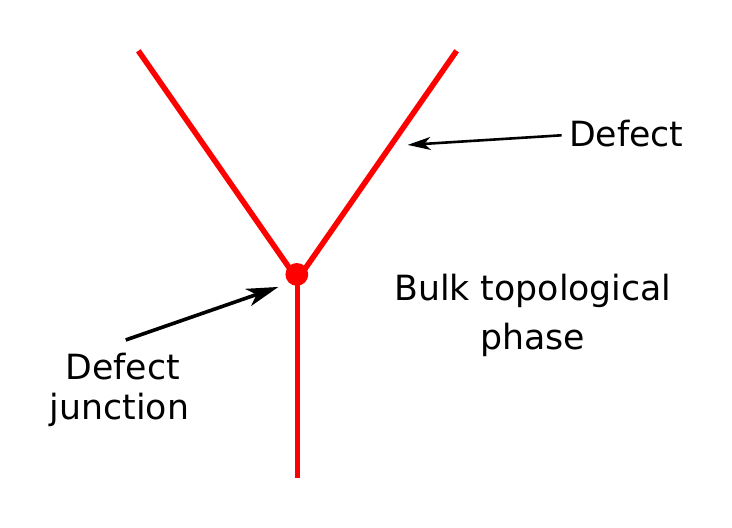}
\caption{\label{fig:defectnetwork}A defect network in 2-D consists of a bulk topological phase, 1-D defects, and 0-D defect junctions (of course, one can also consider the case where the 1-D defects are trivial, in which case a ``0-D defect junction'' is just a point defect). In higher dimensions, one can also have higher-order junctions.}
\end{figure}

Given the competing nature of the two frameworks just described, it is natural to ask whether they are equivalent. In this work, we will unify the two frameworks under the roof of a single mathematical formalism, and thereby answer this question in the affirmative. We will first generalize the ``block state'' picture of Refs.~\onlinecite{SHFH,Huang_1705} to one which works also for non-invertible topological phases, which we call the ``defect network'' picture. (Ref.~\onlinecite{SHFH} also briefly discussed a path to extend ``block states'' to non-invertible phases.) A ``defect network'' consists of a $G$-symmetric network of defects in a $G_{\mathrm{int}}$-symmetric topological phase (where $G_{\mathrm{int}} \leq G$ is the subgroup of internal symmetries). See Figure \ref{fig:defectnetwork} for illustration.

 We show that the defect network picture can be derived in two different ways: first, by physical arguments along the lines of those of Refs.~\onlinecite{SHFH,Huang_1705}; and, second, as a mathematical consequence of our general classification from Ref.~\onlinecite{Thorngren_1612}. This proves the equivalence of the two approaches. In the case of invertible phases, the equivalence between block states and smooth states can be viewed as a manifestation of the mathematical phenomenon of Poincar\'e duality: that is, the isomorphism between generalized homology and cohomology theories on finite-dimensional manifolds.

The outline of our paper is as follows.
In Section \ref{sec:physical}, we discuss the physical picture of defect networks that we are advocating, and motivate it by physical arguments. In Section \ref{sec:smooth_states} we review the notion of smooth states, and explain intuitively why one might expect a classification by smooth states to be equivalent to defect networks. In Section \ref{sec:assumptions}, we review the precise mathematical framework of Ref.~\onlinecite{Thorngren_1612}, which can be viewed as formalizing the notion of smooth states with symmetry. Then, in Section \ref{sec:mathematical} we show rigorously how the defect network picture arises from the mathematical formalism  of Ref.~\onlinecite{Thorngren_1612}. In Section \ref{sec:translations} we discuss defect networks for topological phases in two dimensions with only translation symmetry. In Section \ref{sec:spectral} we outline a mathematical tool, called a ``spectral sequence'', which we expect to be useful for computing properties of defect networks. Finally, in Section \ref{sec:discussion}, we discuss avenues for future investigation and related works.

\section{Defect networks: the physical picture}
\label{sec:physical}

\subsection{Defect networks (physical)}
\label{sec:defect_networks_physical}

\begin{figure}
\includegraphics[width=7cm]{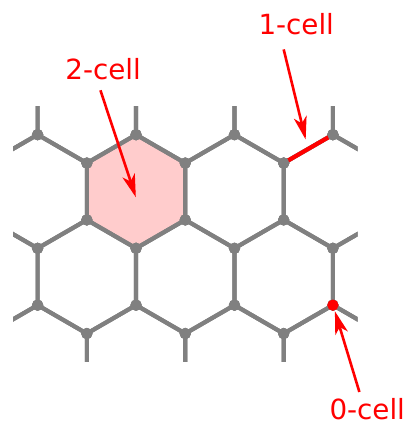}
\caption{\label{cellstructure}A (portion of) a cell decomposition of a 2-dimensional manifold $X$. In the defect network picture, the 2-cells will carry a 2-D topological phase, the 1-cells will carry 1-D defects, and the 0-cells will carry junctions between 1-D defects.}
\end{figure}

A crystalline topological phase exists inside a $d$-dimensional manifold $X$ which represents the physical space in which the system is embedded. We assume that $X$ is acted upon by a symmetry group $G$ (some elements of $G$ can act trivially on $X$, in which case they represent internal symmetries). In most physical cases, we would want to take $X = \mathbb{R}^d$, and the $G$ action to be by isometries of Euclidean space, for example reflections, rotations, translations, glide reflections, and so forth.

The geometrical picture of a crystalline topological phase (both SPT and SET phases) will be an object that we refer to as a \emph{defect network}. For simplicity of exposition, we will define a defect network in terms of a cell decomposition of $X$ (see Figure \ref{cellstructure}), although we expect that similar notions exist in the continuum.
We will choose the cell decomposition such that the image of a cell under the action of any $g \in G$ is itself a cell. Moreover, for each cell $\Sigma$, let $G_\Sigma$ be the subgroup of $G$ that maps $\Sigma$ to itself. We require that each element of $G_\Sigma$ leaves each point in $\Sigma$ fixed. (One can show that if $G$ is a group of isometries of Euclidean space acting on $\mathbb{R}^d$, then a cell decomposition satisfying the required properties always exists. In fact, the minimal such cell decomposition is closely related to the ``Wyckoff positions'' of the space group.)

What we want to imagine is that the cells can be chosen to be very large compared to the lattice spacing and the correlation length. This is easy to do in the case where $G$ is just a point group; if $G$ also contains spatial translations, then this will require that the translation unit cell size $a$ be much greater than the lattice spacing and the correlation length. The restriction to phases that admit ground states satisfying this property underlies the framework both of Ref.~\onlinecite{Thorngren_1612} and of Refs.~\onlinecite{SHFH,Huang_1705}

 First consider the top-dimensional cells, i.e. $d$-cells (in that case, $G_\Sigma$ is just the subgroup of internal symmetries, which we call $G_{\mathrm{int}}$). In the interior of each $d$-cell, we can forget about all the symmetries except the internal ones, and specify the $G_\mathrm{int}$ symmetric topological phase $S_d$ of the system inside the $d$-cell. Then, while respecting the whole $G$ symmetry, it is possible to symmetrically deform the system such that inside each $d$-cell the state of the system is a fixed reference ground-state for the $G_{\mathrm{int}}$-symmetric topological phase on the cell (by choosing a $G_{\mathrm{int}}$-symmetric local unitary on one element of an orbit of cells under $G$, and then using the $G$-related local unitaries on each of the other cells in the orbit\cite{SHFH}).

Next consider a $(d-1)$-cell $\Sigma$. In general, its symmetry group $G_\Sigma$ contains the symmetry group of the $d$-cells which it adjoins, but might be larger. We can identify the state of the system on the $d-1$-cell as being a \emph{$d-1$-dimensional $G_\Sigma$ symmetric defect} between the $d$-dimensional topological phases carried on the adjoining $d$-cells. Such defects can have distinct classes which cannot be deformed into each other. We will choose to deform the state of the system inside of the $d-1$-cell into a fixed reference configuration for the class of defects which it is in (this can always be done $G$-symmetrically by similar arguments to before).

Crucially, in this paper we will only consider a subset of all possible defects, which are sufficient to describe the kind of crystalline topological phases that are classified by the approach of Ref.~\onlinecite{Thorngren_1612} (which were there called ``crystalline topological liquids''). The class of defect we consider is called a \emph{smoothable defect}. A smoothable defect is one which can be implemented in the arena of smooth states as defined in Ref.~\onlinecite{Thorngren_1612} and discussed here in Section \ref{sec:smooth_states}; this means that one can write down a parent Hamiltonian for the defect which varies very slowly as a function of space, such that the state remains gapped at all points in space. Note that this immediately implies that the $d$-dimensional $G_{\mathrm{int}}$ symmetric topological phases carried on top-dimensional cells must be equal, because if they are connected by a smoothable defect then this implies they are connected as a function of Hamiltonian parameters without a phase transition.

What we expect (and will prove in the case where the bulk phase is invertible, but not in the general case) is that a defect $\mathfrak{d}$ is smoothable if and only if it is invertible: that is, the defect does not separate two different bulk phases and, if the symmetries are allowed to be lifted explicily, there is an inverse defect $\overline{\mathfrak{d}}$ such that the fusion $\mathfrak{d} \times \overline{\mathfrak{d}}$ gives a trivial defect \footnote{A defect $1$ is trivial if $1 \times \mathfrak{d}$ is equivalent by a local unitary to $\mathfrak{d}$ for any defect $\mathfrak{d}$}.
 An example of a defect which is explicitly \emph{not} invertible (and we therefore do not consider) would be a 2-D toric code embedded in a 3-D system.

Now let us move on, and consider a $(d-2)$-cell $\Sigma$. We can think of the state of the system on $\Sigma$ as representing a $(d-2)$-dimensional \emph{defect junction} between the $(d-1)$-dimensional defects   on the adjoining $(d-1)$-cells. We can continue in this way until we reach $0$-dimensional higher defect junctions. Moreover, by deforming the state of the system on each $k$-cell to the appropriate reference configuration, we find that up to deformations, the overall system can be specified by the $d$-dimensional topological phase carried on $d$-cells and by the defect class carried on $k$-cells for $0 \leq k < d$. This data specifies what we call a \emph{defect network}.

In order to turn these ideas into a general classification of crystalline topological phases, one needs a general understanding of invertible defects in topological phases (possibly with higher symmetry than the phase itself), which to our knowledge has not yet been developed. In some sense, such a theory will be obtained in this paper, from a mathematical perspective, in the course of deriving the defect network picture from the general framework of Ref.~\onlinecite{Thorngren_1612}.

However, if we specialize to the case of crystalline SPT phases (or, more generally, invertible crystalline topological phases), we can be more concrete. Indeed, it is easy to argue (see, for example Ref.~\onlinecite{SHFH}) that $k$-dimensional invertible $G_\Sigma$-symmetric defects in a $d$-dimensional invertible phase form a torsor over the classification of $k$ invertible topological phases with symmetry $G_\Sigma$ [the torsor becomes a group, i.e. it has a natural identity element, in the case where all the $k'$-dimensional defects are trivial for $k' > k$]. Note that on each $k$-cell $\Sigma$, $G_\Sigma$ is effectively acting as an \emph{internal} symmetry. Thus, we can leverage what we already know about the classification of topological phases with \emph{internal} symmetries to understand crystalline topological phases; this idea was deployed to great effect in  Refs.~\onlinecite{SHFH,Huang_1705}.

\subsection{Anomalies (physical)}
\label{sec:anomalies_physical}
The above arguments demonstrate that we can always (subject to the condition about the correlation length being much less than the lattice spacing) deform any ground state into a canonical ``defect network'' state. So in order to classify crystalline phases we have to characterize defect networks. The idea is that we should first classify the phase on $d$-dimensional cells, then classify defects of dimension $d-1$ in said phase, then for any configuration of dimension-$d-1$ defects on $d-1$-cells, classify the possible junctions on $d-2$-cells, and so forth. Here we want to emphasize a subtlety: the need for there to be an invertible defect junction on $k$-cells places a non-trivial restriction on \emph{which} defect classes are allowed on $r$-cells for $r > k$. In general, we say that an \emph{anomaly} occurs on a $k$-cell $\Sigma$ when there is \emph{no} possible invertible junction between the defect classes on higher-dimensional cells. (In a gapped symmetric defect network state, there should be no anomalies).

\begin{figure}
\includegraphics[width=6cm]{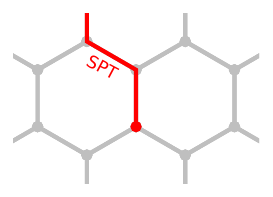}
\caption{\label{fig:termination}A 1-D defect (in this case, a 1-D SPT) inducing an anomaly on a 0-cell}
\end{figure}

\begin{figure}
\includegraphics[width=5cm]{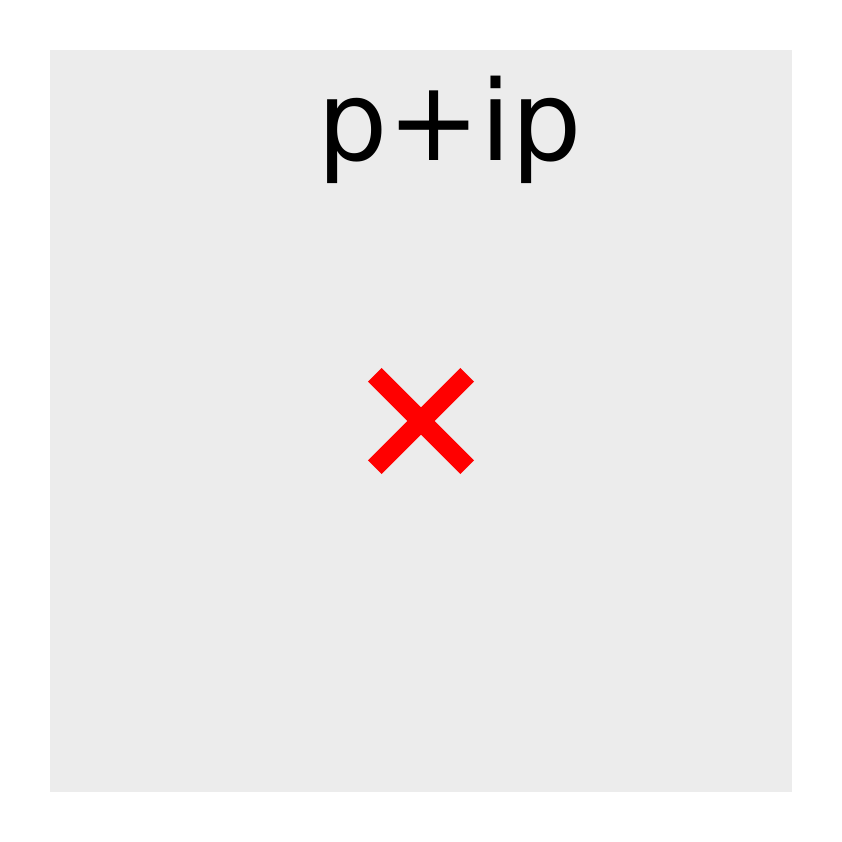}
\caption{\label{fig:ppip}A $C_2$ symmetric $p+ip$ superconductor in 2-D (with $R^2 = 1$) induces an ``anomaly'', i.e. a Majorana zero mode, two dimensions lower.}
\end{figure}

To illustrate this phenomenon, let us restrict ourself to the case of invertible phases, and for simplicity we will assume that the data associated to $r$-dimensional cells is trivial for $r > k_0$. Then the data associated to a $k_0$-cell $\Sigma$ simply a $k_0$-dimensional invertible phase with symmetry $G_\Sigma$. The statement then is that some such data assignments are anomalous. A simple example of anomalous data is shown in Figure \ref{fig:termination}, for the case where the whole symmetry $G$ acts internally. Because, in the configuration shown in Figure \ref{fig:termination}, the 1-D SPT terminates at a point, there must be degenerate edge modes at this point, transforming under a non-trivial projective representation of the symmetry. Therefore, Figure \ref{fig:termination} cannot depict a gapped non-degenerate symmetric ground state, and hence a crystalline SPT.

The above discussion is an example of the general statement that the data associated with $k_0$-cells can lead to an anomaly in dimension $k = k_0-1$. More generally, the anomaly could appear for any $k < k_0$. In such cases, the cause of the anomaly can be more subtle. As an example, consider a fermionic system in $d=2$ with $C_2$ rotation symmetry [with the rotation symmetry satisfying $R^2 = +1$, not $R^2 = (-1)^F$], where the top-dimensional cells carry a $(p+ip)$ superconductor. Now let us try to write down a field theory to describe the long-wavelength limit of this system. It is well known that the $(p+ip)$ superconductor can be described in the continuum by the Hamiltonian
\begin{multline}
H = \int d^2 \mathbf{r} \biggl[ \Psi^{\dagger}\left(-\frac{1}{2m}\nabla^2 - \mu\right) \Psi \\+ \Delta \Psi^{\dagger} (\partial_x + i \partial_y) \Psi^{\dagger} + h.c. \biggr],
\end{multline}
where $\Psi(\mathbf{r})$ is a fermionic field, and $\Delta, m$ and $\mu$ are constants. The problem is that we want the continuum theory to preserve the $C_2$ rotation symmetry, and the pairing term is not rotationally invariant, as can be seen by writing it in polar coordinates $(r,\theta)$:
\begin{equation}
\label{pplusip_pairing}
\int d^2 \mathbf{r} \Psi^{\dagger}(\partial_x + i \partial_y) \Psi^{\dagger}
= \int r dr d\theta \, e^{i\theta} \Psi^{\dagger} (\partial_r + i r \partial_\theta) \Psi^{\dagger}.
\end{equation}
On the other hand, we can make \eqnref{pplusip_pairing} rotationally invariant if we redefine $\Psi^{\dagger} \to e^{i\theta/2} \Psi^{\dagger}$, which removes the $e^{i\theta}$ factor. However, this introduces a new problem: the field redefinition changes the boundary conditions for circling around the origin, thus effectively introducing a $\pi$ vortex (flux of fermion parity) at the origin; in a $p+ip$ superconductor, this binds an emergent Majorana zero mode. Thus, in this case putting $p+ip$ superconductor on 2-cells gives rise to an anomaly two dimensions lower (see Figure \ref{fig:ppip}).

Observe that the two examples above share the feature that when an anomaly appears on a $k$-cell, it is always classified by $h^{k+1}(BG_\sigma)$, where $h^k(BH)$ denotes the classification of invertible phases in $k$ spatial dimensions with internal symmetry $H$. In other words, the anomaly looks like the \emph{boundary} of a $k+1$-dimensional $G_\sigma$-symmetric phase. In Section \ref{sec:anomalies_mathematical}, we will show that in invertible crystalline topological phases the anomalies always take this form.

One consequence of this result is that (at least for invertible crystalline topological phases) the anomaly can always be resolved at the \emph{surface} of a $d+1$-dimensional state with symmetry $G$. Observe that, since the $d+1$-dimensional state by definition admits an invertible gapped boundary while preserving the symmetries, it is necessarily a \emph{trivial} crystalline SPT. For example, the $C_2$ symmetric $p+ip$ superconductor discussed above can occur at the surface of a 3-D state with $C_2$ rotation symmetry which carries a Kitaev chain on the rotation axis. In the next section, we will explain why such a state is a trivial crystalline SPT in 3-D.

Finally, let us note that, in the case of invertible bosonic phases, anomalies on 0-cells [which are  characterized by projective representations of $G_\Sigma$, classified by group cohomology $\mathcal{H}^2(G_\Sigma, U(1))$] can be ``cancelled'' if the \emph{microscopic} degrees of freedom, i.e. those used to define the Hilbert space in which the ground state lives, also carry a projective representation at the corresponding points (one can think of this as a special case of the surface terminations discussed in the previous paragraph). This has interesting consequences for Lieb-Schultz-Mattis type theorems, which will be explored in more detail in a forthcoming work\cite{Else_inpreparation}.

\subsection{Deformations (physical)}
\label{sec:deformations_physical}
Any defect network which is not anomalous as discussed in the previous section represents some allowed state. But we still have to determine which such states cannot be smoothly connected in the presence of the symmetry, i.e.\ what are the \emph{equivalence classes} of defect networks that characterize crystalline topological phases? Therefore, we introduce the notion of a defect network deformation. Although we have so far worked in terms of a cell structure on $X$, deformations are most naturally understood in the continuum. It should be clear how the notion of a defect network generalizes to the continuum: we simply allow $k$-dimensional defects to exist on any $k$-dimensional submanifold, instead of only on the $k$-cells of the cell structure. Then a deformation just means that we allow the configuration of the defects to vary in a smooth way, as long as the spatial symmetry is always respected. We also have to consider the possibility of fusion of defects.

We emphasize that there will in general be fusion moves that relate defects of different dimension. For example, consider a 3-D phase with $C_2$ rotation symmetry, carrying a Kitaev chain on the rotation axis. There is no deformation purely in the space of one-dimensional defects which can trivialize this state. However, one can imagine bringing in a $C_2$ symmetric cylinder surrounding the rotation axis, and carrying a $p+ip$ superconductor. By similar arguments to Section \ref{sec:anomalies_physical}, we find that if we shrink the cylinder to the rotation axis, it will leave behind a Kitaev chain, which can cancels the Kitaev chain that was originally on the rotation axis\cite{Freed_unpublished}.

It might not be obvious that deformations and fusions of defect networks generate all possible deformations of \emph{states} (of which defect networks are just some limit). Let us now show that any deformation of states can be understood in terms of deformations of defect networks. This will also provide a cellular formulation of defect network deformations, which will be useful later on.

 Recall that any deformation of gapped ground states can be understood in terms of the action of a local unitary (LU), i.e. a finite-depth quantum circuit \cite{Hastings2005a,Osborne2007,Chen2010}. Any such local unitary $U$ has a property that we call the ``light cone radius'', which, roughly, is the distance over which quantum information spreads under the action of $U$ (more precisely, for a finite-depth quantum circuit of depth $k$ such that each layer is a product over non-intersecting regions of diameter $l$, the light-cone radius is $kl$). We consider only LUs with light-cone radius that is much smaller than the size of the cells. (If there is no translation symmetry, there is no restriction on the light-cone radius).  Otherwise, if we tried to interpret the LU as a deformation of gapped ground states, it would pass through intermediate states which violate the condition that the correlation length should be much less than the size of the cells. We call an LU satisfying this condition a \emph{cellular local unitary}, or cLU.

 We note that allowing non-cellular LUs in our equivalence relation might logically decrease the number of phases in the classification (because phases previously considered distinct could be related by a non-cellular LU). However, for any space group $G$ with a translation symmetry, there are subgroups $H < G$ isomorphic to $G$ with arbitrarily large unit cell. Our classification, as derived based on the cellular LU equivalence relation, has the property that for some infinite subset of these subgroups, any pair of phases which are distinct with $G$ symmetry are also distinct with $H$ symmetry. We demonstrate this in Appendix \ref{s:bigunit}. Thus, if we have a (possibly non-cellular) LU circuit $U$ which maps between two $G$-symmetric states, then since the light-cone radius is finite, there is an $H < G$ where $U$ can be made cLU for an $H$-invariant coarse graining of the cell structure; hence they are equivalent $H$-phases according to our classification, so by the property mentioned above they are also equivalent $G$-phases in our classification. It follows that allowing non-cellular cLU does not change the number of phases after all.

We say that a cLU is a $k$-cLU if it acts only in the vicinity of the $k$-skeleton, i.e.\ the union of the cells of dimension $k$. We say that a $k$-cLU is a strict $k$-cLU if acts trivially near the $k-1$-skeleton. One can show that any $k$-cLU can be written as a product $U_k = U_{k-1} V$, where $U_{k-1}$ is a $(k-1)$-cLU and $V$ is a strict $k$-cLU. We say two states are $k$-equivalent if they can be related by a $k$-cLU.

\begin{figure}
\includegraphics[width=7cm]{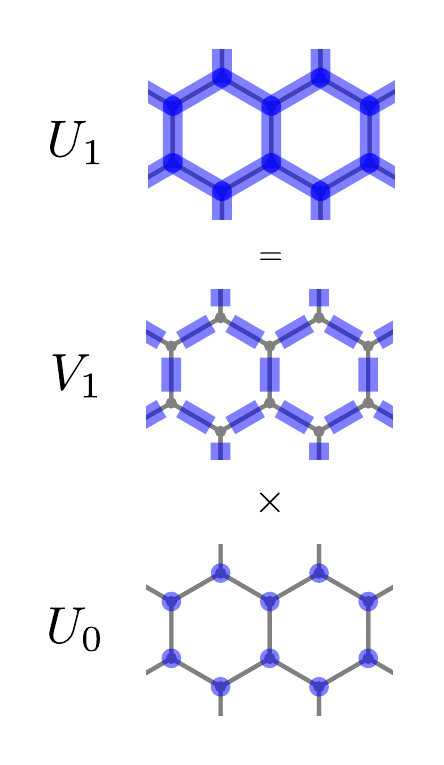}
\caption{\label{fig:unitary}A $k$-cLU can be written as the product of a strict $k$-cLU and a $k-1$-cLU}.
\end{figure}

\begin{figure}
\includegraphics[width=4cm]{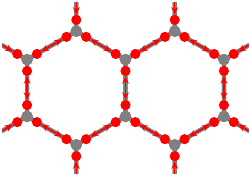}
\caption{\label{fig:unitary_4} The effect of $V_k$ is to create new defects near the boundary of each $k$-cell. In the langauge of defect network deformations, we can think of these as having been created out of the vacuum along each $k$-cell. The newly created excitations then fuse onto $k-1$-cells. In this illustration, $k=1$.}
\end{figure}

Now consider a defect network $C_0$, and suppose a defect network $C$ is $k$-equivalent to $C_0$. Then there exists a $k$-LU $U_k$ such that $U_k \ket{C_0} = \ket{C}$, where $\ket{C}$ and $\ket{C_0}$ are the corresponding states. We can always $U_k$ as $U_{k-1} V$, where $U_{k-1}$ is a $(k-1)$-cLU and $V_k$ is a strict $k$-cLU (see Figure \ref{fig:unitary}). Thus, $V_k \ket{C} = U_{k-1}^{\dagger} \ket{C_0}$. Note that $U_{k-1}^{\dagger} \ket{C_0}$ looks the same as $\ket{C_0}$ inside of $k$-cells, i.e. inside of $k$-cells it still looks like a canonical representative of a defect class. Moreover, we know from the fact that $C$ and $C_0$ are $k$-equivalent that they must have the same defect class on $k$-cells, and in particular (since $\ket{C}$ is a defect network) it must look like the same canonical representative on $k$-cells. In other words, acting with $V_k$ has no effect inside of $k$-cells. So the only possible effect is to create $k-1$-dimensional defects near the edge of the $k$-cells. Therefore, we interpret the equation $\ket{C} = U_{k-1} V_k \ket{C_0}$ as saying that we create $k-1$-dimensional defects near the edge of $k$-cells and then fuse them onto $k$-cells to create a defect network state (see Figure \ref{fig:unitary_4}). Note that for a given $V_k$ (that is, a given pattern of defects created on $k$-cells) and a given $C_0$, there may still be several different defect networks $C$ that can be created by such a process, according to different ways of doing the fusion at $k-1$-cells. For example, suppose that $U_{k-1} V_k \ket{C_0}$ and $U_{k-1}' V_k \ket{C_0}$ are both defect network states, which we call $\ket{C}$ and $\ket{C'}$. Then we see that that $\ket{C}$ and $\ket{C'}$ are $k-1$-equivalent, because $\ket{C} = (U_{k-1}')^{-1} U_{k-1} \ket{C'}$. Hence, we can conclude inductively that the process of creating defects and fusing them as described indeed generates all possible deformations between defect networks.

\section{The ``dual'' physical picture: smooth states}
\label{sec:smooth_states}
\begin{figure*}
    \subfloat[][Smooth state]{\includegraphics[scale=0.5]{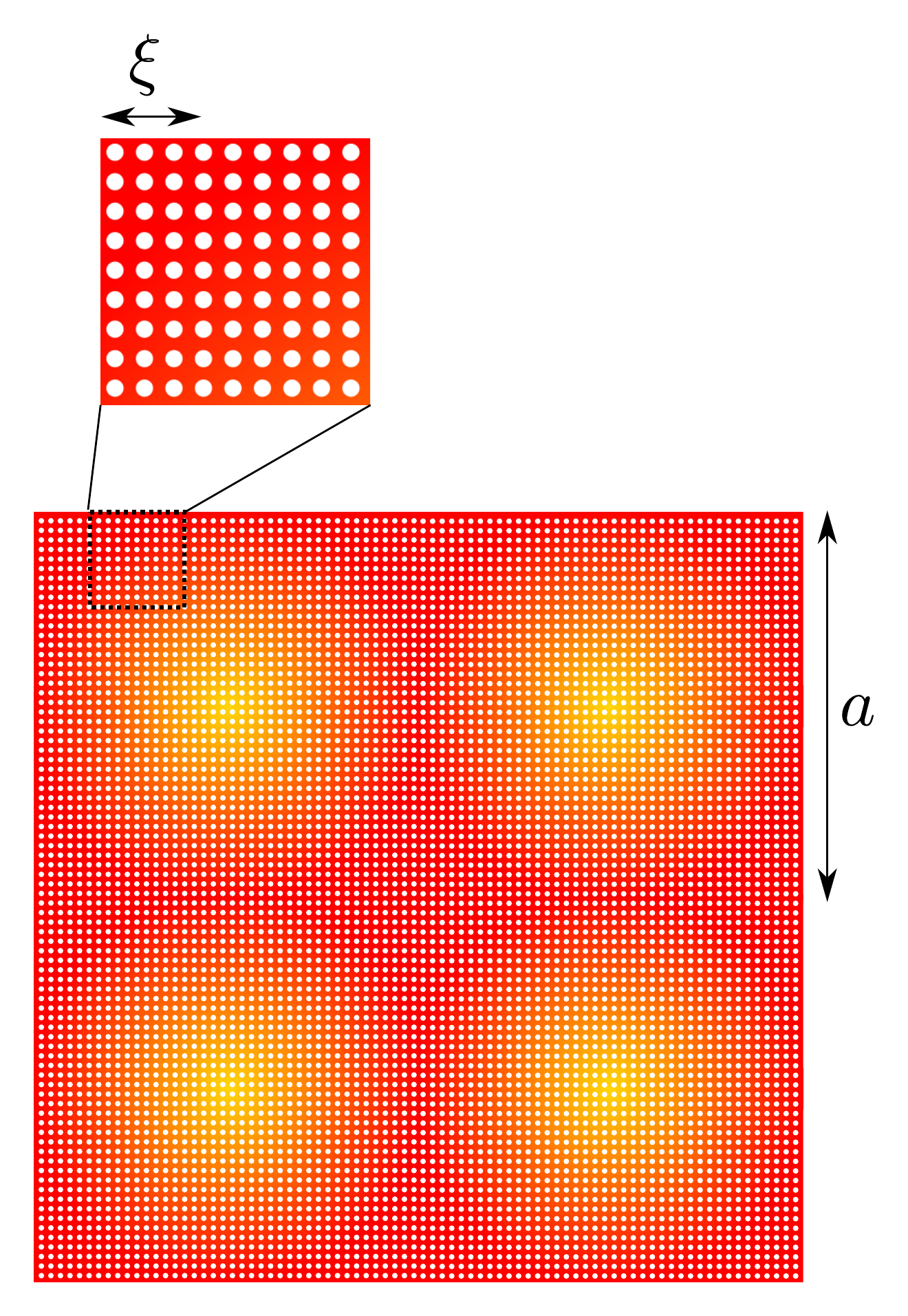}}
\subfloat[Forgetting about microscopic details]{\includegraphics[scale=0.5]{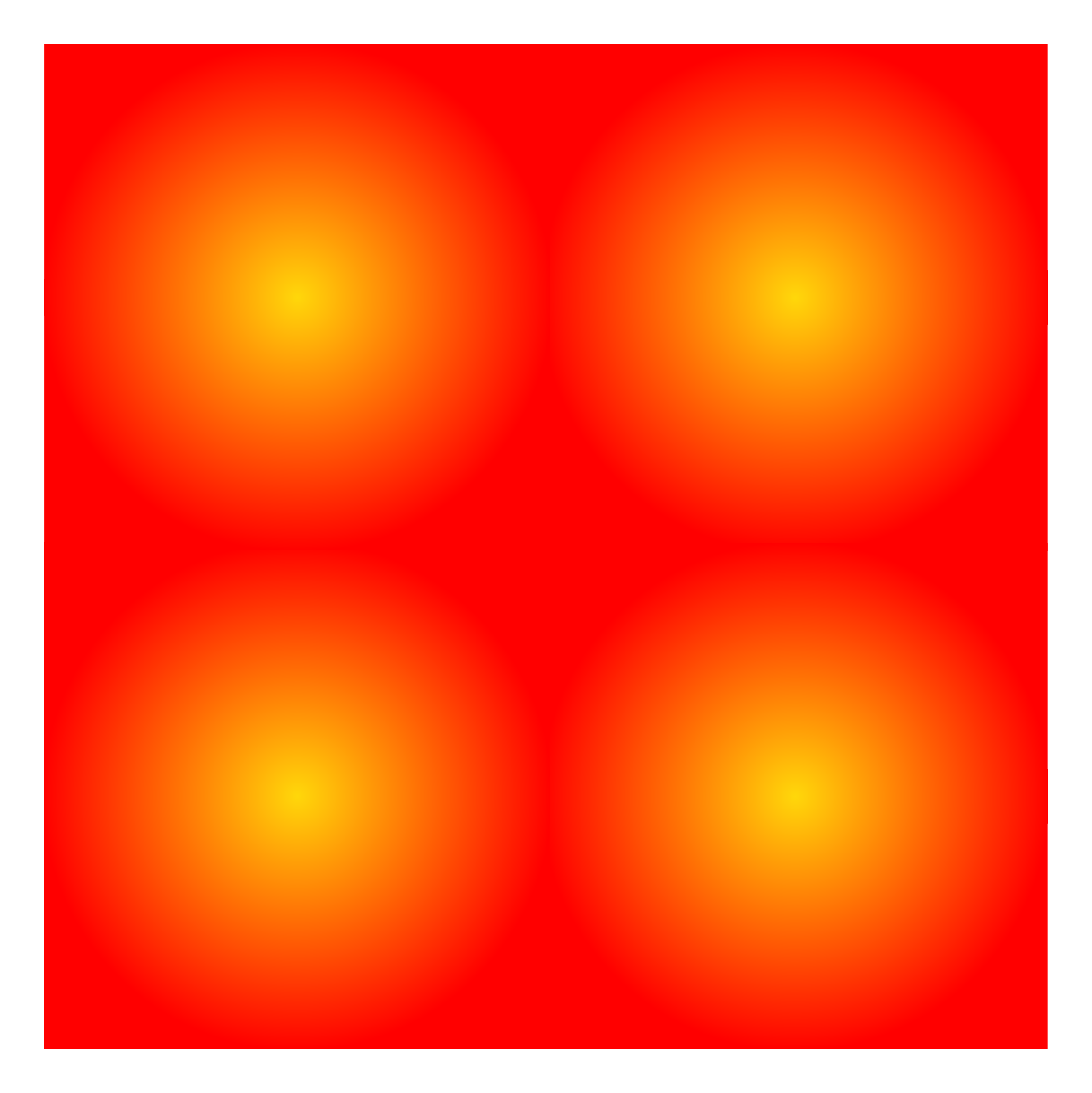}}
\end{figure*}
In this section we will review a different picture of crystalline topological phase in terms of smooth states, as previously introduced in Ref.~\onlinecite{Thorngren_1612}. One of the goals of this paper is that show that the two pictures from this section and the previous section are actually equivalent.

A smooth state is supposed to represent a particular kind of physical state. As in the previous section, we assume that the correlation length $\xi$ and microscopic lattice spacing $a_0$ are much less than the translation unit cell size $a$. In contrast to the previous section, we assume that on scales much less than some radius of variation $R$ (which is much larger than $\xi$ and $a_0$) there is an \emph{approximate} translation symmetry (we emphasize that this is distinct from the translation subgroup of the spatial symmetry $G$, which is exact if it is present). That is, on scales small compared with $R$, the state varies only very slowly with space.

Now the idea is to assume that the details of the lattice at the microscopic scale are not very important, and so we can ``abstract out'' and define a smooth state on a spatial manifold $X$ to be a map
\begin{equation}
\label{smooth_state_defn}
f : X \to \Theta_d
\end{equation}
for some space $\Theta_d$ which is an abstraction of the $d$-dimensional states in the neighborhood of a given point. One (albeit very abstract) way\cite{Thorngren_1612} to think of $\Theta_d$ is as the space of all topological quantum field theories, where a point in the space $\Theta_d$ is a TQFT, a continuous path in $\Theta_d$ is an isomorphism between TQFTs, a deformation between paths is an equivalence between isomorphisms, and so forth.
 We can also implement internal or spatial symmetries in smooth states by requiring the map $f$ to be $G$-equivariant (we leave the precise mathematical formulation to the next section).

 Let us note that, in general, if the tangent bundle of $X$ is non-trivial, we should think of the space of local states at every point as forming a non-trivial fiber bundle over $X$ with fiber $\Theta_d$; in that case we replace \eqnref{smooth_state_defn} with a section of this bundle (for details, see Appendix \ref{sec:twists}).

 Now we must ask why the ``smooth state'' picture should be equivalent to the ``defect network'' picture. We will give the detailed argument later on, but the idea is basically to ``sharpen'' the smooth state by deforming it, concentrating its spatial variation, so that near any $k$-cell in the cell decomposition of $X$ discussed in the previous section, the smooth state is approximately constant in the directions tangent to the cell. This is describing a $k$-dimensional  defect localized near the cell. Since the defect is obtained from sharpening a smooth state, it is obviously smoothable in the sense defined in Section \ref{sec:defect_networks_physical}. (Note that, in some sense, this is just applying the arguments of Section \ref{sec:defect_networks_physical} at the level of smooth states).

 \begin{figure}
 \includegraphics[width=5cm]{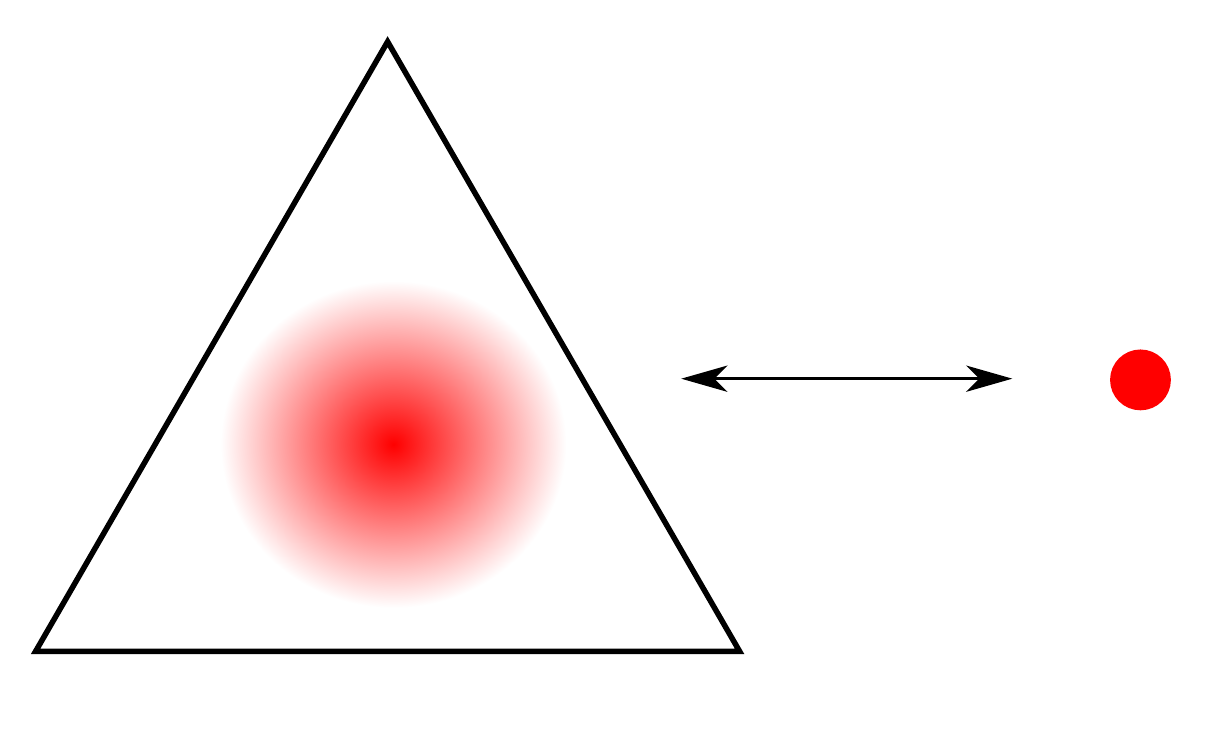}
 \caption{\label{duality}The duality principle for smooth states. Smooth states on a manifold with boundary, of dimension $n+k$, which can be contracted to a manifold $M$ of dimension $k$, are in one-to-one correspondence with smooth states on $M$. In this picture, $k=0$ and $n=2$.}
 \end{figure}
 Finally, in the case of invertible phases, there is one additional idea involved in the relationship between defect networks and smooth states. Recall that, physically a $k$-dimensional invertible defect in a \emph{trivial} phase is supposed to be equivalent to a $k$-dimensional invertible phase (from which it follows that $k$-dimensional defects in a non-trivial invertible phase are a torsor over $k$-dimensional invertible phases). Since we claimed (so far, without proof) that an \emph{invertible} defect is the same as a \emph{smoothable} defect, it follows that smooth states on a $k+r$-dimensional tubular neighborhood $N$ of an $k$-dimensional manifold $M$, constrained so that the local state on the boundary of $N$ is trivial, are in one-to-one correspondence with smooth states on $M$ (for example, see Figure \ref{duality}). This should hold even if there are spatial symmetries acting on $N$ that leave $M$ fixed; in that case, on $M$ they act as \emph{internal} symmetries.
 We call this the \emph{duality principle}.
 This implies highly non-trivial relations between the spaces $\Theta_k$ for different $k$; the mathematical formulation is the ``generalized cohomology'' assumption discussed in the next section.

\section{Implementing symmetries in a smooth state: general classification of crystalline phases}
\label{sec:assumptions}
Now let us discuss how to implement the symmetries in a smooth state; this will allow us to recover the general classification of Ref.~\onlinecite{Thorngren_1612}. Our starting point is the following conjecture

\begin{conjecture}
The classification of topological phases (SPT or SET) in $d$ dimensions with \emph{internal} symmetry $G$ is
given by homotopy classes of maps
\begin{equation}
f : BG \to \Theta_d,
\end{equation}
 Here $BG$ is the
so-called ``classifying space'' of a group $G$; up to homotopy
equivalence, it is specified as $BG = EG/G$, where $EG$ is any
contractible space with a free action of $G$.
\end{conjecture}
This conjecture has previously appeared in various forms \cite{Kitaev2006,Freed_1406,KitaevIPAM,Thorngren_1612,Xiong_1701,Gaiotto_1712}.

Ref.~\onlinecite{Thorngren_1612} proposed how to generalize this to describe crystalline topological phases, in a way that also extends the notion of smooth states defined in the previous section (by showing how to implement symmetries in a smooth state):

\begin{conjecture}
     The classification of topological phases in $d$ dimensions (SPT or SET) with \emph{spatial}
        symmetry $G$ which acts on (physical) space $X$ (usually we would want to
        take $X = \mathbb{R}^d$) is given by homotopy classes of maps
        \begin{equation}
            f : X // G \to \Theta_d
        \end{equation}
        where $\Theta_d$ is the \emph{same} space as in Conjecture 1. Here
        $X//G$ denotes the ``homotopy quotient'' of $X$ by the action of $G$.
        Up to homotopy equivalence, this is specified as $X // G = (X \times EG)/G$, where
        $EG$, as before, is a contractible space with a free action of $G$, and $G$
        acts diagonally on the product space $X \times EG$.
\end{conjecture}
Conjectures 1 and 2 have an immediate corollary, which Ref.~\onlinecite{Thorngren_1612} called the ``Crystalline Equivalence Principle''. In the case where $X = \mathbb{R}^d$ (or, generally, $X$ is any
contractible space), it is a mathematical fact that $X//G$ and $BG$ are homotopy
equivalent. (To see this note, just observe that in this case $X \times EG$ is
itself a contractible space with a free action of $G$). Thus, one immediately concludes

\begin{corollary}[Crystalline Equivalence Principle]
The classification of topological phases with internal symmetry $G$ is the \emph{same} as the classification of topological phases with \emph{spatial} symmetry $G$.
\end{corollary}

Let us note that for systems of fermions, and bosonic systems with orientation-reversing symmetries, Conjecture 2 must be slightly modified (even if $X = \mathbb{R}^d$), as we discuss in Appendix \ref{sec:twists}; then the map $f$ becomes a section of a fiber bundle over $X//G$ with fiber $\Theta_d$. (This point of view also accounts for the ``twists'' in the Crystalline Equivalence Principle that we discussed in Ref.~\onlinecite{Thorngren_1612}, such as the orientation-reversing symmetries mapping to anti-unitary symmetries).
For simplicity, in the main body of the paper we will assume that the bundle is trivial and Conjecture 2 holds as written, but the arguments can easily be extended to the general case.

Although the Crystalline Equivalence Principle can be a useful way to compute the classification mathematically, it can be difficult to physically interpret the resulting phases, since (a) the connection between the topological phase with spatial symmetry and the corresponding topological phase with internal symmetry is often obscure; and (b) generally spatial symmetry groups $G$ are quite large, and non-Abelian, making the interpretation of the topological phase with internal symmetry $G$ a challenge in itself.

In this work, we intend to address this issue by showing that crystalline phases classified according to Conjecture 2 are described by defect networks.

\subsection{Invertible phases and generalized cohomology}
We will often want to restrict ourself to the case of invertible crystalline topological phases. In this case, we will need to make an additional assumption \cite{KitaevIPAM,Xiong_1701,Gaiotto_1712,FH}:
\begin{conjecture}
 For invertible phases, the spaces $\Theta_d$ appearing in Conjectures 1 and 2 can be taken to satisfy
\begin{equation}
\Theta_d \simeq \Omega \Theta_{d+1},
\end{equation}
where ``$\simeq$'' denotes homotopy equivalence, and $\Omega \Theta_{d+1}$ is the based loop space of $\Theta_{d+1}$, i.e.\ the set of maps $\gamma : [0,1] \to \Theta_{d+1}$ such that $\gamma(0) = \gamma(1) = \vartheta_*$ for a fixed basepoint $\vartheta_* \in \Theta_{d+1}$ (which is supposed to represent the trivial ``vacuum'' state).
\end{conjecture}
In mathematical terms, this is saying that the spaces $\Theta_{\bullet}$ form an ``$\Omega$-spectrum''. Physically, it is the statement of the ``duality principle'' mentioned in Section \ref{sec:smooth_states}, for systems without spatial symmetries (it turns out that it also implies the duality principle in the presence of spatial symmetries, but this is non-trivial to show; see Appendix \ref{sec:twists}).
Equivalently, it is saying that the classification of invertible phases is $d$ space dimensions with internal symmetry $G$ can be expressed as $h^{d}(BG)$ [or $h^{d}(X//G)$ in the spatial case], where $h^{\bullet}(-)$ is a ``generalized cohomology theory''.

Let us note that a wide variety of proposed partial classifications for invertible interacting phases of bosons or fermions satisfy this property. Examples include the ``group cohomology'' \cite{CGLW} and ``cobordism'' \cite{K} classifications of bosonic SPTs, and the ``group supercohomology'' \cite{Gu2014} and ``spin cobordism'' \cite{Gaiotto2016, Kapustin_1701} classifications of fermionic SPTs. It also holds for the Freed-Hopkins classification of invertible topological quantum field theories \cite{FH}. Therefore, our results will hold with respect to all such classifications.

\section{The mathematical picture}
\label{sec:mathematical}
In this section, we will show how the phenomena discussed from a physical point of view in Section \ref{sec:physical} can be recovered through rigorous mathematical arguments, given the assumptions of Section \ref{sec:assumptions}. Note that the subsections of this section will exactly parallel those of Section \ref{sec:physical}.

\subsection{Defect networks (mathematical)}
\label{sec:defect_networks_mathematical}
Let us show how the defect network picture can be obtained from the general considerations of Section \ref{sec:assumptions}. The argument will be expressed in terms of the map $f : X // G \to \Theta_d$ posited in Conjecture 2.
 The reader will note that the arguments here, though couched in mathematical language, look structurally \emph{very} similar to the more physical arguments we used to justify the defect network picture in Section \ref{sec:defect_networks_physical}. Indeed, this is a reflection of the fact that the map $f$ can be interpreted as specifying a kind of ground state, namely a smooth state.

\begin{figure}
\includegraphics[width=7cm]{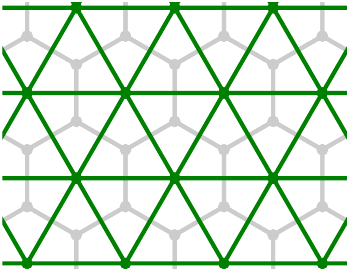}
\caption{\label{honeycomb_dual} The dual (shown in green) of the honeycomb cell decomposition previously shown in Figure \ref{cellstructure}. In general, there is a one-to-one correspondence between $k$-cells in the original cell decomposition and $d-k$-cells in the dual cell decomposition.}
\end{figure}

Our arguments will be expressed in terms of the \emph{dual} cell decomposition of the one on which the defect networks live, e.g.\ as shown in Figure \ref{honeycomb_dual}. There is a one-to-one correspondence between $k$-cells $\Sigma$ and dual $d-k$-cells $\widetilde{\Sigma}$. Moreover, $\Sigma$ and $\widetilde{\Sigma}$ intersect at a single point, the barycenter $x_\Sigma$. The subgroup $G_\Sigma \leq G$ that maps a cell $\Sigma$ to itself is the same subgroup that maps the dual cell $\widetilde{\Sigma}$ to itself. Note, however, that whereas we chose the original cell $\Sigma$ to have the property that $G_x = G_\Sigma$ for all $x$ in the interior of $\Sigma$, the barycenter $x_\Sigma$ is the only point in $\widetilde{\Sigma}$ that necessarily has $G_{x} = G_{\Sigma}$.

 Finally, let us note that for the purpose of these arguments we will assume that $X$ is a $d$-dimensional Riemannian manifold and the action of $G$ on $X$ is metric-preserving. (Obviously, this is true if $X = \mathbb{R}^d$ and $G$ acts on $X$ by Euclidean isometries, as is the case for the space groups one normally considers in physics.) This implies that the action of $G$ on $\widetilde{\Sigma}$ is linear and orthogonal; that is, a dual $k$-cell $\widetilde{\Sigma}$ can be identified with a subset of $\mathbb{R}^k$ such that the action on $\widetilde{\Sigma}$ is induced by a representation of $G$ in the orthogonal group $\mathrm{O}(k)$.

Now recall that, according to Conjecture 2, a crystalline topological phase corresponds to a map $f : X//G \to \Theta_d$, for some space $\Theta_d$. In this section, whenever we form the homotopy quotient $Y//H$ with respect to any space $Y$ and any subgroup $H \leq G$, we will mean $Y//H = (Y \times EG)/H$ (we are allowed to use $EG$ here, because it is a contractible space on which $H$ acts freely).

On the dual cellulation, we can represent $f$ by the following data

\begin{itemize}
\item
To each dual $k$-cell $\wSigma$, we associate a map $f_{\wSigma} : \wSigma // G_\Sigma \to \Theta_d$.
\end{itemize}

This data must satisfy certain consistency relations if it is indeed to describe a map $f : X//G \to \Theta_d$:
\begin{enumerate}[(i)]
\item \label{restriction_condn} Consider a dual $k$-cell $\wSigma$ and a dual $k-1$-cell $\wsigma$ which is a face of $\wSigma$. Then there is an obvious map $\varphi : \wsigma // G_\sigma \to \wSigma // G_\wSigma$. We require that $f_\wSigma \circ \varphi = f_\wsigma$.
\item \label{equivariance_condn} For any dual $k$-cell $\wSigma$ and any $g \in G$, there is an obvious homeomorphism $\varphi :  \wSigma // G_\Sigma \to (g\wSigma) // (G_{g\sigma})$, and we require that $f_{g\wSigma} \circ \varphi = f_{\wSigma}$.
\end{enumerate}

\begin{lemma}\label{cellular_map}The space of collections of maps $f_\Sigma$ satisfying the above conditions is equivalent (i.e. homeomorphic) to the space of maps $f : X // G \to \Theta_d$.

\begin{proof} A function $f : X // G \to \Theta_d$ is equivalent to a $G$-invariant function $\hat{f} : X \times EG \to \Theta_d$, and a function $f_\wSigma : \wSigma // G_\Sigma \to \Theta_d$ is equivalent to a $G_\Sigma$ invariant function $\hat{f}_\wSigma : \wSigma \times EG \to \Theta_d$. In terms of these maps, condition (\ref{restriction_condn}) amounts to saying that the restriction of $\hat{f}_\wSigma$ to $\wsigma \times EG$ is equal to $\hat{f}_\wsigma$, and condition (\ref{equivariance_condn}) amounts to saying that $f_{g\wSigma}(gx,ge) = f_\wSigma(x,e)$ for all $x \in \wSigma, g \in G, e \in EG$. The functions $\hat{f}$ and $\hat{f}_\Sigma$ are then related according to
\begin{equation}
\hat{f}(x,e) = \hat{f}_\wSigma(x,e),
\end{equation}
where $\Sigma$ is any cell containing $x$.
\end{proof}
\end{lemma}

The goal now is to assign a physical interpretation to the maps $f_\wSigma$. As a warm up, let us start with dual $0$-cells, i.e.\ $\wSigma$ is a point (corresponding to a top-dimension, i.e. $d$-cell, $\Sigma$ in the original cell complex). Then if $p$ is a dual $0$-cell, then we have a function $f_p : p // G_p \to \Theta_d$. In fact, $p // G_p$ is homotopy equivalent to $BG_p$. But recall that a map $BG_p \to \Theta_d$ classifies topological phases in $d$ dimensions with internal symmetry $G_p$. The interpretation should be clear: inside of a $d$-cell in the original cell structure, the ``effective'' internal symmetry (subgroup of $G$ that leaves points fixed inside the $d$-cell in the original cell decomposition) is $G_p$, so we can have a $G_p$-symmetric topological phase.

Next, we want to claim that the homotopy classes of maps $f_\wSigma$ on dual $k$-cells $\wSigma$ describe $d-k$-dimensional defect junctions on the original $(d-k)$-cells. The idea is to proceed inductively. After we have characterized the homotopy classes of dual $k-1$-cells as defect junctions, we will deform the associated maps to fixed reference configurations for said defect junctions. Then, for any dual $k$-cell $\widetilde{\Sigma}$, the restriction of the map $f_\wSigma : \wSigma // G_\Sigma$ to $\partial \wSigma // G_\Sigma$ (where $\partial \wSigma$ denotes the $k-1$-dimensional boundary of the dual $k$-cell $\wSigma$) is already completely determined (to see this, invoke Lemma \ref{cellular_map} with the replacement $X \to \partial \wSigma$). Therefore, we must consider homotopy classes of maps $f : \wSigma//G_\wSigma \to G_\wSigma$ whose restriction to $\partial \wSigma // G_\wSigma$ is held fixed.

We assert that such homotopy classes on a dual $k$-cells should precisely be identified with classes of smoothable $G_\Sigma$-symmetric defects junctions on the original $(d-k)$-cells, where the junctions are formed at the intersection of the junctions on the original $d-k+1$-cells. One way to see this is by applying the ``spatially dependent TQFT'' idea from Ref.~\onlinecite{Thorngren_1612} to the classification of (smoothable) defects. However, let us discuss two cases in which this assertion can be seen more straightforwardly.

Firstly, we can consider the case in which $G_\wSigma$ leaves all the points in a dual $k$-cell $\wSigma$ (in terms of the original $d-k$-cell $\Sigma$, this is saying that $G_\Sigma$ is the same as $G_{\Sigma}'$ for any $d-k+1$-cell $\Sigma'$ of which $\Sigma$ forms part of the boundary; that is, $\Sigma$ has the same symmetry as its surroundings). In that case, $\wSigma // G_\wSigma = \wSigma \times BG_\Sigma$. So we have a map $f_\wSigma : \Sigma \times BG_\wSigma \to \Theta_d$, which we know is supposed to restrict to a fixed map $f_\wsigma : \wsigma \times BG_\wSigma \to \Theta_d$ on any face $\wsigma$ (note that $G_\Sigma$ acting trivially on $\Sigma$ implies that $G_\Sigma = G_\sigma$). So we effectively have a map $B^k \times BG_\Sigma \to \Theta_d$, where $B^k$ is the $k$-ball, with the restriction to the boundary of the $k$-ball held fixed. How should we interpret this map? An answer is supplied by interpreting $\Theta_d$ as the space of TQFTs\cite{Thorngren_1612}. In the context of TQFTs such maps are well-understood to describe invertible codimension-k defect junctions in topological phases with an internal $G_\Sigma$ symmetry\cite{KICM,KS,FRS,Henriques}.

The second case to consider is that of invertible crystalline topological phases. Recall that in this case, one can argue physically that the classes of codimension-$k$ defect junctions living on a $(d-k)$-cell $\Sigma$ should form a torsor over the invertible topological phases with internal symmetry $G_\Sigma$. We want to show that this is what we obtain from the homotopy classes of maps $f_\wSigma$ on dual $k$-cells $\wSigma$. Indeed, this follows from the following Lemma (setting $H = G_\Sigma$, $r = d$, and noting using the fact that the $G_\Sigma$ action on $\widetilde{\Sigma}$ is supposed to be linear orthogonal):

\begin{lemma}
\label{thom_lemma}
Let $H$ be a group with linear orthogonal action on the $k$-ball $B^k$. Then
homotopy classes of maps $f : B^k \to \Theta_\bullet$ with \emph{fixed} restriction to $\partial B^k // H$ (where $\partial B^k$ is the boundary of $B^k$) are a torsor over homotopy classes of maps $f : BH \to \Theta_{\bullet-k}$, with a natural identity element in the case where the fixed restriction is the constant map.
\begin{proof}
Note that, strictly speaking, this is not the precise statement of the Lemma; to make it precise we have to use the more general definition of smooth states as sections of a bundle with fiber $\Theta_d$ (as mentioned earlier). The result then follows from the \emph{Thom isomorphism} of generalized cohomology. For the details, see Appendix \ref{sec:twists}.
\end{proof}
\end{lemma}

\subsection{Anomalies (mathematical)}
\label{sec:anomalies_mathematical}

Let us discuss the mathematical interpretation of the anomalies discussed in Section \ref{sec:anomalies_physical}. The basic idea is as follows. Suppose that we have a map $f_{k_0} : X_{k_0} // G \to \Theta_d$, where $X_{k_0}$ is the $k_0$-skeleton of dual cells (i.e.\ the union of all dual $r$-cells for $r \leq k_0$). Applying the arguments of the previous section shows that it can be characterized up to homotopy by $d-k_0$-dimensional defect junctions on $k_0$-cells. Now the question is whether such a map $f_{k_0}$ can be extended to a map $f : X // G \to \Theta_d$ defined on the full space. In general, this will not be possible, and this will correspond to an anomaly.

Specifically, what can happen is that there is an obstruction to consistently extending the maps $f_{\widetilde{\Sigma}}$ on dual $k_0$-cells $\widetilde{\Sigma}$ to a map $f_\sigma$ on some dual $k$-cell $\widetilde{\sigma}$ containing $\widetilde{\Sigma}$ (for some $k > k_0$). In the case of invertible phases, one can show that this obstruction is valued in
$h^{d-k+1}(BG_\sigma)$. To see this, note if that $\widetilde{\sigma}$ is the first cell on which the obstruction appears, then it must have been possible to extend consistently to its boundary $\partial \widetilde{\sigma}$ at least. Then we invoke the fact
that the inclusion $(\partial \widetilde{\sigma}) // G_\sigma \to \widetilde{\sigma} // G_\sigma$ induces a long exact sequence in generalized cohomology, of which a portion looks like:
\begin{multline}
\cdots \to h^{d}(\widetilde{\sigma} // G) \to h^{d}(\partial \widetilde{\sigma} // G) \\ \to h^{d+1}( \widetilde{\sigma} // G; (\partial\widetilde{\sigma}) // G) \to \cdots
\end{multline}
Remember that for any space $S$, $h^d(S)$ computes the homotopy classes of maps $f : S \to \Theta_d$. Therefore, this exact sequence is telling us that the obstruction to extending a map $(\partial \widetilde{\sigma})//G$ to $\widetilde{\sigma}//G$ is valued in the relative cohomology $h^{d+1}( \widetilde{\sigma} // G; (\partial\widetilde{\sigma}) // G)$. Then Lemma \ref{thom_lemma} tells us that this object is isomorphic to $h^{d-k+1}(BG_\sigma)$.

\subsection{Deformations (mathematical)}
\label{sec:deformations_mathematical}
To understand deformations mathematically, we follow an argument with a similar structure to the physical argument from Section \ref{sec:anomalies_physical}, with the differences coming from the fact that we are now working with the dual cells. Let $f, f' : X//G \to \Theta_d$ be two maps which have been deformed to the canonical form on each cell, as discussed in Section \ref{sec:defect_networks_mathematical}. We say say that a homotopy $\hat{f} : [0,1] \times X//G \to \Theta_d$, such that $\hat{f}(0,\cdot) = f$ and $\hat{f}(1, \cdot) = f'$, is a $k$-homotopy if it is the constant homotopy when restricted to the the $k-1$ skeleton of the dual cells.

Consider a $k$-homotopy $\hat{f}$. Then for any dual $k$-cell $\widetilde{\Sigma}$, we obtain a map $\hat{f}_{\widetilde{\Sigma}}$ into
$\Theta_d$ from $[0,1] \times \widetilde{\Sigma}//G_\Sigma = (\widetilde{\Sigma} \times [0,1])//G_\Sigma$ (where we define $G$ to act trivially on $[0,1]$).
Observe that, on the surface of $\widetilde{\Sigma} \times [0,1]$, $\hat{f}$ is completely constrained by $f_{\widetilde{\Sigma}}$ and $f_{\widetilde{\Sigma}}'$.
Recall that we postulated in Section \ref{sec:defect_networks_mathematical} (and this can be shown more explicitly for invertible phases, given the generalized cohomology hypothesis) that for $k$-cells $\widetilde{\Sigma}$ with fixed restriction to their boundary, the maps $\widetilde{\Sigma} // G_\Sigma \to \Theta_d$ correspond to $d-k$-dimensional defect junctions. We can treat $\widetilde{\Sigma} \times [0,1]$ itself as $k+1$-cell, so the map $\hat{f}$ should correspond to a $d-k-1$-dimensional defect boundary. We interpret this as saying that a $d-k-1$-dimensional defect is getting pumped to the boundary of the (original, not dual) $d-k$-cell $\Sigma$, as discussed physically in Section \ref{sec:deformations_physical}. In general, this pumping data on $k$-cells corresponds to the restriction $\hat{f}_k$ of $\hat{f}$ to the $k$-skeleton of the dual cells. Two homotopies $\hat{f}$, $\hat{f}'$ with the same restriction $\hat{f}_k$ can be related by composition with a $k+1$-homotopy, so inductively we conclude that the pumping on cells of all dimension completely characterizes  homotopies. Note that we can in principle derive the fusion rules for defects from the requirement that $\hat{f}$ be non-anomalous on $([0,1] \times X)//G$ in the sense described in Section \ref{sec:anomalies_mathematical}.

\section{Defect networks in two dimensions with only translation symmetry}
\label{sec:translations}
\begin{figure}
\includegraphics[width=8cm]{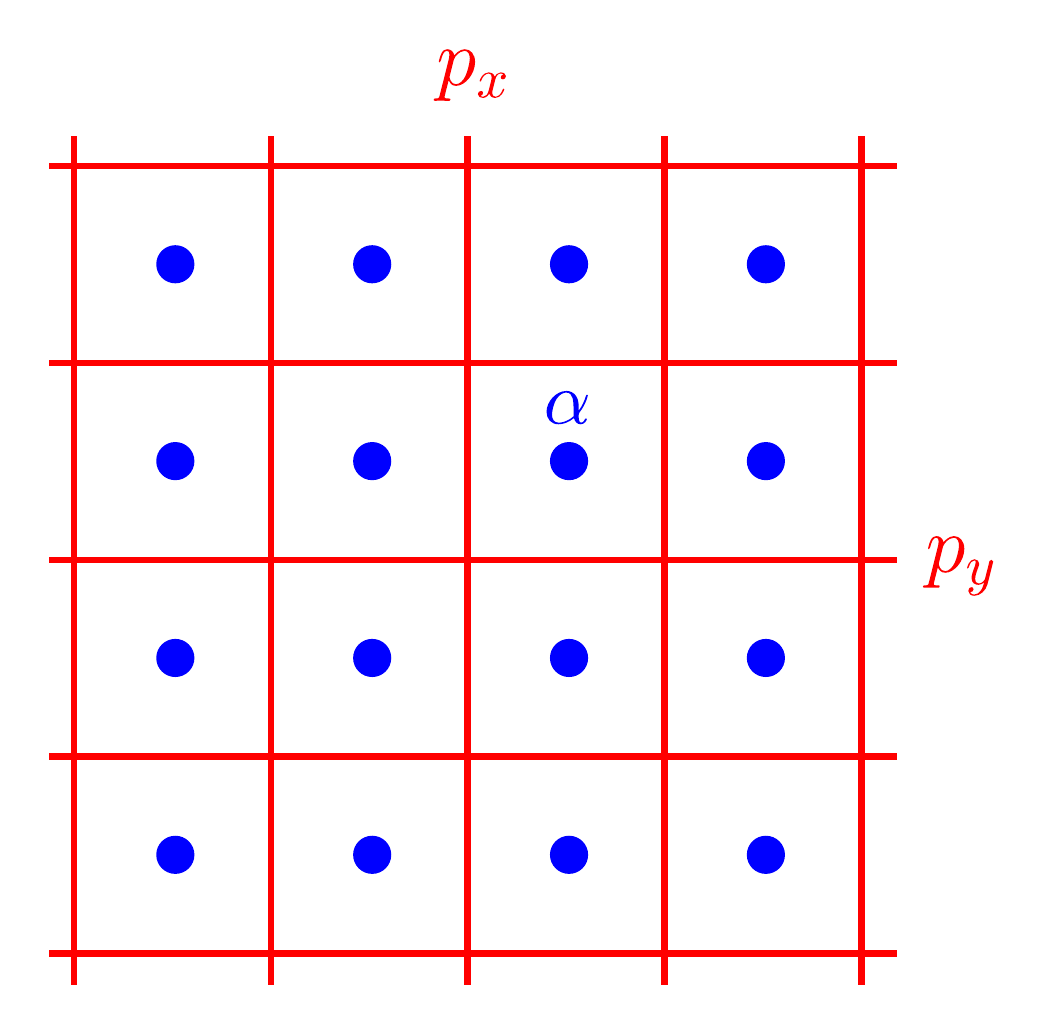}
\caption{\label{fig:lines}The data for a defect network in two diensions with translation symmetry. $p_x$ and $p_y$ are duality twist lines, and $\alpha$ is an Abelian anyon type.}
\end{figure}
The main purpose of this paper is to show that defect networks in principle reproduce all the physics of crystalline topological phases. We will not, however, go very much into how this works in concrete examples. ``Block states'', which are the specialization of defect networks to invertible phases, have been studied quite systematically in Refs.~\onlinecite{SHFH,Huang_1705}. Here we will content ourselves with discussing a few simple examples of defect networks for non-invertible phases. Specifically, we will consider the case where the bulk phase (i.e.\ the topological phase when the symmetries are lifted) is a 2-dimensional bosonic topologically ordered phase supporting anyonic excitations, and the only symmetry present is discrete translation symmetry in the $x$ and $y$ directions.

Let us first consider the case where the only symmetries are translations, $G = \mathbb{Z} \times \mathbb{Z}$. We can work out what the classification of symmetry-enriched phases should be by invoking the Crystalline Equivalence Principle, and then using the classification of SET phases with internal symmetry from Refs.~\onlinecite{BBCW,ENO}. We find that the phases should be classified by a group homomorphism $\rho : \mathbb{Z} \times \mathbb{Z} \to \mathrm{Aut}(\mathcal{C})$, where $\mathrm{Aut}(\mathcal{C})$ is the group of permutations of the anyon labels that leave the braiding statistics unchanged, and by a symmetry fractionalization class $[\omega] \in \mathcal{H}^2_{\rho}(\mathbb{Z} \times \mathbb{Z}, A) \cong A / A_\rho$, where $A$ is the Abelian group of Abelian anyons, $A_{\rho}$ is the subgroup generated by  $\{ (g \cdot a - a) : a \in A, g \in G \}$, and $G$ acts on $A$ according to the its image by $\rho$.

Now let us see how to understand this classification in terms of defect networks. Firstly, observe that the homomorphism $\rho$ is uniquely determined given its images $p_x = \rho(T_x), p_y = \rho(T_y)$, where $T_x$ and $T_y$ are the translation generators. To each $p \in \mathrm{Aut}(\mathcal{C})$, there is the notion of a ``duality twist line'', such that particles passing through the twist line are acted upon by the permutation $p$. So we can consider a spatial arrangement of twist lines as shown in Figure \ref{fig:lines}, with vertical and horizontal lines corresponding to the permutations $p_x$ and $p_y$ respectively. Note that, if translation symmetry is preserved, we can shift the twist lines in space, as long as they all move together, but never remove them or change their character. Therefore, each $p_x$,$p_y$ defines a distinct defect network.

To understand the class $[\omega]$, we note that an anyon is a dimension-0 defect (an invertible defect only if the anyon is Abelian), and therefore we can define an invertible defect network where each unit cell contains some anyon $\alpha \in A$. The only way to change the anyon type carried per unit cell is if some duality twist lines are present, in which case we can create an anyon $\beta$ , along with its anti-particle $-\beta$, out of the vacuum in each unit cell, and then move each $\beta$ over a duality twist line. This sends $\alpha \to \alpha + p(\beta) - \beta$, where $p = p_x$ or $p_y$, which explains why we obtain an $A/A_\rho$ classification.

\section{Spectral sequences}
\label{sec:spectral}
The arguments of this paper are sufficient to demonstrate, both conceptually and rigorously, the equivalence of the defect network picture and the general classification of Ref.~\onlinecite{Thorngren_1612}. \emph{In principle} the arguments we have given can be used to determine, for example, the fusion rules for defects and the possible anomaly associated with each defect network. On the other hand, we have not yet developed tools to allow one to compute such things \emph{in practice}. There are two different approaches one could envision: firstly, one could attempt an analysis in particular cases on purely physical grounds, as was done by Refs.~\onlinecite{SHFH,Huang_1705}, and trust that this should reproduce the same result as the general mathematical framework. However, it might also be desirable to do the computation in the mathematical framework directly. Here we briefly describe what we expect to be the key tool, at least for invertible phases, namely a \emph{spectral sequence}; we leave the details for future work. In the special case of the group cohomology classification of bosonic SPT phases, more detailed computations can be found in Ref.~\onlinecite{Zhida_1810}.

In mathematics, a spectral sequence takes the form of a sequence of pages $E^r$. Each such page can be written as a two-dimensional array of Abelian groups, which we write as $E^r_{p,q}$ (where $E^r_{p,q} = 0$ unless $p,q \geq 0$). There are homomorphisms $d^r$ (the ``differentials'') that act on each page according to
\begin{equation}
d_r : E^r_{p,q} \to E^r_{p+r, q-r+1}
\end{equation}
Moreover, the differentials satisfy $d_{r+1} \cdot d_r = 0$, and the $r+1$-th page can be computed from the $r$-th page according to
\begin{equation}
E^{r+1} = \ker{d_r}/\operatorname{im}{d_{r-1}}.
\end{equation}
In general, such a spectral sequence always converges: that is, for each $p,q$, for large enough $k$ the differentials will have source and target spaces that are outside the first quadrant (i.e. not $p,q \geq 0$), in which case
\begin{equation}
 E^{k+1}_{p,q} \cong E^k_{p,q} := E^{\infty}_{p,q}.
\end{equation}

Let  $\mathcal{C}^d := h^{d}_G(X)$ be the full classification group for invertible crystalline phases on a $d$-dimensional manifold $X$ with $G$-action, taking into account the twists (required for bosonic systems with orientation-reversing symmetries or fermionic systems, as described in Appendix \ref{sec:twists}). Define $\mathcal{C}_k$ to be the subgroup of $\mathcal{C}_d$ describing phases which can be realized as a defect network containing defects of dimension $\leq k$. What we will do is to construct a spectral sequence such that
\begin{equation}
\label{blockstateconvergence}
E^{\infty}_{k+1, d-k} = \mathcal{C}_k/\mathcal{C}_{k-1}.
\end{equation}
Therefore, from the spectral sequence one can recover $\mathcal{C}_d$, up to an extension problem of Abelian groups.

In fact, the individual pages and differentials of the spectral sequence have a very physical interpretation. The $E^r_{k+1,d-k}$ entry describes an ``approximation'' to $\mathcal{C}_k/\mathcal{C}_{k-1}$ taking into account only deformations of defects of dimension at most $k+r$, and only anomalies of dimension at least $k-r$. Thus, one can compute $E^{r+1}$ from $E^r$ by taking into account an extra dimension of deformations (described by the incoming differential $d_r$ at $E^r_{k+1,d-k}$) and an extra dimension of anomalies (described by the outgoing differential).

The construction of the spectral sequence in our case proceeds by considering the series of inclusions (a ``filtration'')
\begin{equation}
X_0 \subseteq X_1 \subseteq \cdots \subseteq X_{d-1} \subseteq X,
\end{equation}
where $X_k$ is the $k$-skeleton of the dual cell structure on $X$. One can show that when the classification is governed by a generalized cohomology theory $h^{\bullet}(-)$ (that is, the spaces $\Theta_{\bullet}$ form an $\Omega$-spectrum), as we assumed to be the case for invertible phases, then this filtration
 induces a spectral sequence (for example, see Ref.~\onlinecite{Kono__04} for the untwisted case) $E^{r}_{p,q}$ such that
\begin{equation}
E^{\infty}_{p,q} = F^p h^{p+q}_G(X) / F^{p+1}h^{p+1}_G(X),
\end{equation}
where
\begin{equation}
F^p h^{\bullet}(X // G) = \ker\left( i_p : h^{\bullet}_G(X_p) \to h^{\bullet}_G(X_{p-1}) \right),
\end{equation}
This indeed implies \eqnref{blockstateconvergence}. Moreover, the first page of the spectral sequence is given by the (twisted) relative cohomology
\begin{equation}
\label{relative_cohomology}
E^1_{p,q} = h^{p+q}_G(X_p, X_{p-1})
\end{equation}
In terms of the maps $f : X // G  \to \Theta_{\bullet}$ introduced in Section \ref{sec:defect_networks_mathematical}, the ``relative cohomology'' \eqnref{relative_cohomology} means the homotopy classes of maps $ X_p // G \to \Theta^{p+q}$ which restrict to the constant map on $X_{p-1}//G$ (or the analogous statement in the twisted case).

Using the methods of Section \ref{sec:defect_networks_mathematical}, we see that $E^{1}_{k+1,d-k}$ contains precisely the data associated with the (original, not dual) $k$-cells in a defect network. To see this, note if we start following the general approach of Section \ref{sec:defect_networks_mathematical}, but replacing $X$ with $X_{d-k}$, then we find that the only data we need are the maps $f_\wSigma : \widetilde{\Sigma}//G_\Sigma$ associated to dual $d-k$-cells $\wSigma$, and these are all constrained to be constant on their boundaries by assumption. Then Lemma \ref{thom_lemma} shows that the homotopy classes of such maps are simply classified by $h^{d-k}(BG_\Sigma)$.

The $E^1$ page does not ``know'' about the anomalies and deformations described in Sections \ref{sec:anomalies_mathematical} and \ref{sec:deformations_mathematical}. However, these get taken into account in higher pages of the spectral sequence. When the generalized cohomology theory $h^{\bullet}(-)$ is just ordinary cohomology, then this spectral sequence reduces to the one considered in more detail in Ref.~\onlinecite{Zhida_1810}.

\section{Discussion}
\label{sec:discussion}
In this work, we have demonstrated a general picture for understanding topological phases with spatial symmetries and show how it agrees with previously proposed frameworks. We hope that it will allow for a better physical understanding of such phases, especially once one moves beyond the formal and general aspects the theory, as developed here, to consider more concrete examples. Indeed, the ``block state'' picture for invertible phases, which is a special case of our picture, has been applied in a variety of cases to give physical pictures of crystalline topological phases\cite{SHFH,Huang_1705,Song_1810}. (As one example, it allows for a very transparent understanding of so-called ``higher-order'' phases which carry gapless modes on lower-dimensional submanifolds of the boundary \cite{Rasmussen_1809}). It would no doubt be fruitful to perform similar analyses for the more general ``defect network'' picture described here for non-invertible phases.

Another avenue of inquiry would be to consider potential generalizations of defect networks. In particular, in this work we considered only ``smoothable'' defects (which we conjectured to be equivalent to invertible defects). It would be interesting to determine what kind of phases can be described using networks of \emph{non}-smoothable defects. For non-smoothable defects, the topological phase carried on top-dimensional cells does not need to be equal on different cells, so in general these defects will be boundaries between different top-dimensional topological phases, and then junctions between such boundaries, and so forth. Moreover in the non-smoothable case, the defect networks might not even need any spatial symmetries for protection. We speculate\cite{AasenCommunication} that the phases of matter describable in this way will be precisely the so-called \emph{fracton phases}
\cite{Chamon_0404,Bravyi_1006,Castelnovo_1108,
     Haah_1101,Bravyi_1112,Yoshida_1302,Vijay_1505,Vijay_1603,
     Williamson_1603,Shirley_1712,Nandkishore_1803}.
     This would be an intriguing connection between fracton phases and crystalline topological phases.

\subsection{Related works}
In the course of preparing this manuscript, we became aware of several related preprints \cite{Song_1810,Zhida_1810,Shiozaki_1810}. Unlike any of these works, we also discussed non-invertible phases. Let us make some comments on each preprint, and its relation to our work, in turn.
\begin{itemize}
\item In Ref.~\onlinecite{Song_1810}, the authors use a block states (there named ``topological crystals'') approach to classify \emph{non-interacting} phases of fermions. (Recall that for interacting phases, block states are the specialization of defect networks to invertible phases.) The arguments used have overlap with the physical arguments we presented in Section \ref{sec:physical}. Whether the mathematical derivation of Section \ref{sec:mathematical} will apply depends on whether our assumptions from Section \ref{sec:assumptions}, namely Conjectures 1, 2 and 3 hold for the non-interacting classification, which is not immediately clear.

\item In Ref.~\onlinecite{Zhida_1810}, the authors discuss very systematically a picture for invertible phases which is equivalent to the specialization of our defect network picture to invertible phases. Moreover, for the special case where the internal SPT classification is assumed to be group cohomology, they derive their picture from our general framework of Ref.~\onlinecite{Thorngren_1612} through a spectral sequence. This spectral sequence is a special case of the one we discuss in Section \ref{sec:spectral}.

\item In Ref.~\onlinecite{Shiozaki_1810}, the authors show based on certain assumptions that the ``block state'' picture (that is, the special case of ``defect networks" for invertible phases) can be derived through a spectral sequence. The assumption of Ref.~\onlinecite{Shiozaki_1810} is that the classification of crystalline phases is a ``generalized Bredon equivariant homology theory", a terminology which we adopt, though it is not used in Ref.~\onlinecite{Shiozaki_1810}, to distinguish it from the notions of generalized equivariant (co)homology used by us in this work and in Ref.~\onlinecite{Thorngren_1612}, which we can call ``generalized Borel equivariant (co)homology". Note that there are many different generalized Bredon equivariant homology theories (corresponding to many different choices of ``equivariant spectra''), and Ref.~\onlinecite{Shiozaki_1810} does not attempt to say which one \emph{actually} classifies crystalline SPT phases, except in the case of free fermions (which prevents them from performing any explicit computations). By contrast, our approach is in a sense uniquely determined by the internal SPT classification; specifically, once the classification of SPT phases with \emph{internal} symmetry group $G$ has been identified as a generalized cohomology theory $h^{\bullet}(BG)$, then all the structure of crystalline SPTs can be obtained from $h^{\bullet}$.

There remains, however, the question of whether our classification is a special case of a generalized Bredon homology theory, in which case the arguments of Ref.~\onlinecite{Shiozaki_1810} could be applied to our approach as a special case. Indeed, it is easy to show that our classification is an example of a ``generalized Bredon \textbf{co}homology theory'', defined by replacing the axioms discussed in Ref.~\onlinecite{Shiozaki_1810} by the appropriate cohomology versions. It seems plausible that when the space $X$ which the system physically inhabits is a finite-dimensional manifold, there should be some kind of Poincar\'e duality theorem that relates our classification to a generalized Bredon homology theory.  In fact, for the case where the generalized cohomology theory is ordinary cohomology (that is, we are discussing the group cohomology classification of bosonic SPTs), this can be straightforwardly demonstrated\cite{MO_2}. In general we do not have a proof, but we note that our approach to derive the defect network picture, involving as it does passing to the dual cellulation and invoking the Thom isomorphism, is already highly reminiscent of Poincar\'e duality.

\end{itemize}

\begin{acknowledgments}
	We thank David Aasen, Meng Cheng, Chen Fang, Lukasz Fidkowski, S{\o}ren Galatius, Michael Hermele, Sheng-Jie Huang, Alex Takeda, and Yang Qi for helpful discussions. We also thank Mark Grant and Oscar Randal-Wiliams for helpful answers on MathOverflow\cite{MO_1,MO_2}. DVE was supported by the Microsoft Corporation and by the Gordon and Betty Moore Foundation. RT was supported by the National Science Foundation and the Zuckerman STEM Leadership Fellowship.
\end{acknowledgments}

\appendix

\section{General treatments of twists and the Thom isomorphism}
\label{sec:twists}
\subsubsection{The twisted state bundle}
The bundle of local states over an $n$-dimensional manifold $X$ forms a fiber bundle $\mathfrak{S}_X \to X$ with fiber $\Theta_n$. Let us be more precise about how this bundle is constructed. The idea is that there should be a continuous action of $O(n)$ on $\Theta_n$.
 Indeed, $\Theta_n$ is supposed to represent some approximation to the space of ground states on $\mathbb{R}^n$, so we should be able to act on this space by rotations or reflections. More formally, if $\Theta_{\bullet}$ is chosen to be an $\Omega$-spectrum (as we assumed in Section \ref{sec:assumptions} for invertible phases) or more generally, if $\Theta_n$ is chosen to be the space of $n$-dimensional TQFTs, then the existence of such an $O(n)$ action follows from the cobordism hypothesis\cite{BaezDolan} as proven by Lurie \cite{Lurie}. (In general, the cobordism hypothesis only guarantees an up-to-homotopy action of $O(n)$; to simplify the discussion we will ignore this subtlety in what follows, but the results should still hold).

 In the $\Omega$-spectrum case, one can furthermore show that the $O(n)$ action is compatible with the spectrum structure in the sense that, for $\bullet \geq n$, the equivalence
 \begin{equation}
 \label{orthogonal_spectrum}
 \mathrm{Hom}_*(S^n, \Theta_\bullet) \to \Theta_{\bullet-n}
 \end{equation}
[which is guaranteed to exist by the definition of $\Omega$-spectrum, where $\mathrm{Hom}_*(S^n, \Theta_\bullet)$ is space of based maps from $S^k$ to $\Theta_\bullet$; that is, the maps which send the basepoint of $S^k$ to the vacuum state in $\Theta_n$]
is invariant with respect to the diagonal action of $O(n)$ on the source and target of $\mathrm{Hom}_*(S^n, \Theta_{\bullet})$, where $O(n)$ acts on $\Theta_{\bullet}$ through the inclusion $O(n) \to O(\bullet)$.

Then, for any $n$-dimensional vector bundle $E \to B$ (of which the tangent bundle of a manifold is a special case), whose orthonormal frame bundle we write $O(E) \to B$, we can define the associated state bundle which is the bundle over $B$ whose fibers are
 \begin{equation}
 \label{statebundle}
 \mathfrak{S}_n(E)_b = \mathrm{Hom}_{O(n)}(O(E)_b, \Theta_n),
 \end{equation}
 that is, the space of $O(n)$-equivariant maps from $O(E)_b$, the space of frames at $b$, into $\Theta_n$. The fiber $\mathfrak{S}_n(E)_b$ is equivalent to $\Theta_n$ (but not canonically).

If we consider the case where $B$ is an $n$-dimensional manifold $X$, and $TX$ is its tangent bundle, then we define a \emph{smooth state} on $X$ to be a section $X \to \mathfrak{S}_n(TX)$. Then the interpretation of \eqnref{statebundle} is that we need to specify a set of coordinate axes (i.e.\ a frame) near any given point in order to be able to identify the local state in the vicinity of a given point with the ``canonical'' space of states $\Theta_n$.

Next, if we consider the case of an $n$-dimensional manifold $X$ with smooth action of a symmetry group $G$. Then the action of $G$ on the tangent bundle $T(X)$ induces an $n$-dimensional vector bundle $T_GX \to X // G$, where $T_GX = X//G$.
 Defining the associated state bundle $\mathfrak{S}_n(T_GX)$ to $E$ as before, we can then define an \emph{equivariant smoooth state} to be a section $X // G \to \mathfrak{S}_n(T_GX)$.

 For the connection between equivariant smooth states so defined, and the classification of crystalline phases in terms of ``crystalline gauge fields'' proposed by us in Ref.~\onlinecite{Thorngren_1612}, see Appendix \ref{app:cobordism}. In this section, for simplicity we considered only the case that there are no "internal" twists; that is, there are no anti-unitary symmetries, and (for fermionic phases) the internal symmetries have no non-trivial extension by fermion parity, and spatial symmetries have the extension induced by their spatial action [for example, a $C_2$ rotation generator $R$ satisfies $R^2 = (-1)^F$]. However, in Appendix \ref{app:cobordism} we consider also the more general case.

 \subsubsection{The Thom isomorphism}
Now we are ready to give the proof of the Thom isomorphism, i.e.\ Lemma \ref{thom_lemma}. We specialize to the case where $\Theta_{\bullet}$ is an $\Omega$-spectrum.
Let $E \to B$ be an $n$-dimensional vector bundle. We can define the corresponding sphere bundle $\mathrm{Sph}(E)$ by one-point-compactifying each fiber of $E$ by adding a point at infinity. Meanwhile, we define the state bundle $\mathfrak{S}_{\bullet}(E) \to B$ according to \eqnref{statebundle}.
Define $P^{\bullet}(E) := \underline{\mathrm{Hom}}_{*}(\mathrm{Sph}(E), \mathfrak{S}_E^{\bullet})$, where we introduced the notation that for two bundles $\mathcal{E} \to B$ and $\mathcal{E}' \to B$, $\underline{\mathrm{Hom}}_*(\mathcal{E}, \mathcal{E}')$ is the bundle over $B$ whose fiber at $b$ are given by $\mathrm{Hom}_*(\mathcal{E}_b, \mathcal{E}'_b)$, the space of continuous basepoint-preserving maps from $\mathcal{E}_b$ to $\mathcal{E}'_b$ (we take the basepoint of a fiber of $\mathrm{Sph}(E)$ to be the point added at infinity, and the basepoint of a fiber of $\mathfrak{S}_{\bullet}(E)$ to be the vacuum state).
\begin{lemma}
\label{bundle_lemma}
$P^{\bullet}(E)$ is isomorphic to the trivial bundle $\underline{\Theta_{\bullet-n}} = \Theta_{\bullet-n} \times B$.
\begin{proof}
Let us construct a homeomorphism between the fibers at a point $b \in B$. Then we will first construct a map $f_b : \mathrm{Hom}_*(\mathrm{Sph}(E)_b, \mathfrak{S}_{\bullet}(E)_b) \to \mathrm{Hom}_*(S^n, \Theta{\bullet})$. Then we can compose with \eqnref{orthogonal_spectrum} to get a map $f_b' : \mathrm{Hom}_*(\mathrm{Sph}(E)_b, \mathfrak{S}(E)_b)) \to \Theta_{\bullet-n}$.

Recall that that $\mathfrak{S}_{\bullet}(E)_b = \mathrm{Hom}_{O(n)}(O(E)_b, \Theta_n)$. We can canonically write $O(E)_b$ as the space of orthogonal maps $u : E_b \to \mathbb{R}^n$, or equivalently the space of induced homeomorphisms $\mathrm{Sph}(E)_b \to S^n$. Hence, we can construct a canonical homeomorphism $f_b : \mathrm{Hom}(\mathrm{Sph}(E)_b, \mathfrak{S}_{\bullet}(E)_b) \to \mathrm{Hom}(S^n, \Theta_{\bullet})$ according to
\begin{equation}
f_b(\alpha)(s) = \alpha(u_*^{-1}(s))(u_*),
\end{equation}
for some fixed choice of map $u_* : \mathrm{Sph}(E)_b \to S^n$; when we compose with \eqnref{orthogonal_spectrum}, the map $f_b'$ turns out not to depend on the choice of $u_*$, as a consequence of the $O(n)$ equivariance of \eqnref{orthogonal_spectrum}. One can show that this map on fibers induces a bundle isomorphism between $P^{\bullet}(E)$ and $\underline{\Theta_{\bullet-n}}$.
\end{proof}
\end{lemma}

Now let us consider the space $X = \mathbb{R}^n$ with an orthogonal linear action of $G$, that is, a homomorphism $\varphi : G \to O(n)$. We can consider the equivariant tangent bundle $T_GX= TX // G \to X // G$, and we form the associated state bundle $\mathfrak{S}_{\bullet}(T_GX) \to X // G$. Recall that an equivariant smooth state on $X$ is defined to be a section of this bundle. Let $h^{\bullet}_G(X, \infty)$ be the homotopy classes of sections of the bundle $\mathfrak{S}_{\bullet}(T_GX)$ which can be extended to the one-point compactification of $X$ such that the map $BG \to \Theta_{\bullet-n}$ obtained at the point at infinity is the trivial map. Let $h_G^{\bullet - n}(pt) := h^{\bullet-n}(BG)$ be the homotopy classes of maps $BG \to \Theta_{\bullet - n}$. Then we have

\begin{lemma}[Equivariant Thom isomorphism]
\begin{equation}
h_G^{\bullet}(X,\infty) \cong h_G^{\bullet - n}(pt).
\end{equation}
\begin{proof}
In this case, we can check that the bundle $\mathfrak{S}_{\bullet}(T_GX) \to X//G$ is isomorphic to the bundle $(X \times \Theta_\bullet)//G \to X//G$. Hence, a section of this bundle is a map $X // G \to (X \times \Theta_\bullet)//G$ which must compose with the projection $(X \times \Theta_\bullet)//G \to X//G$ to give the identity map. These are equivalent to maps $X // G \to \Theta_\bullet//G$ by composing with the other projection. On the other hand, we can also treat $X//G$ as a vector bundle over $BG$. Then one can show that the associated state bundle $\mathfrak{S}_{\bullet}(X//G)$ is isomorphic to the bundle $\Theta_\bullet//G \to BG$. It follows that $h^{\bullet}_G(X,\infty)$ exactly corresponds to homotopy classes of sections of the bundle $P^{\bullet}(X//G)$ defined above, with $E = X//G$ and $B = BG$.
 Then we invoke Lemma \ref{bundle_lemma}.
 \end{proof}
\end{lemma}

This immediately gives Lemma \ref{thom_lemma} (reformulated in terms of sections of the state bundle) because the additive structure of $\Theta_{\bullet}$ (coming from the fact that each $\Theta_{\bullet}$ is a loop space, which has a notion of loop composition) ensures that is sufficient to prove Lemma \ref{thom_lemma} for the case where the restriction to the boundary of the $k$-ball is trivial, and in that case we can collapse the boundary to a point.

Finally, let us briefly note the form which these results take in the case where the generalized cohomology theory under consideration is
 cobordism\cite{K} (for bosons) or spin cobordism\cite{Gaiotto2016, Kapustin_1701} (for fermions), which are the best current candidates for the ``correct'' classification of SPT phases. In that case, we have\cite{Thorngren_1612} that $h_G^n(X,Y) = \Omega^n_{str}(X//G, Y//G, TX \oplus \xi)$, where $str$ refers to oriented cobordism for bosons and spin cobordism for fermions, and $TX \oplus \xi$ is the twisting bundle, with $TX$ encoding the spatial twist and $\xi$ the internal twist\cite{KTTW}. In this case, for $Y \subset X$ a $G$-equivariant codimension $k$ submanifold, $S(Y)$ the fiber-wise one-point compactification of the normal bundle of $Y$ (a $k$-sphere bundle over $Y$), we have by the usual equivariant Thom isomorphism,
\[\Omega^{n-k}_{str}(Y//G,TY \oplus \xi) = \Omega^n_{str}(S(Y)//G,Y,TS(Y) \oplus \xi),\]
since $TS(Y)|_Y = NY \oplus TY$. Taking $Y$ to be a point yields the above lemma.

\section{Crystalline Topological Liquids and the Baez-Dolan-Lurie Cobordism Hypothesis}
\label{app:cobordism}

In an earlier paper of ours\cite{Thorngren_1612}, given a target manifold $X$ with a smooth action of a group $G$ and a representation $\xi$ of $G$ (equivalently a vector bundle over $BG$) we defined a bosonic (fermionic) $\xi$-twisted crystalline gauge field on a spacetime $M$ as a map $f:M \to X//G$ together with an orientation (spin structure) on $TM \oplus f^*TX//G \oplus \alpha^* \xi$, where $\alpha:M \to BG$ is obtained by composition of $f$ with the projection $X//G \to BG$.

The collection of such $n$-manifolds $(M,f)$ can be described in terms of a cobordism $n$-category of the usual type described eg. by Lurie \cite{Lurie}. These cobordism categories are for manifolds with $(P,\rho)$ structure, where $P$ is a space with an $\mathbb{R}^n$ bundle $\rho$, and a $(P,\rho)$ structure on an $n-k$ manifold $M$ with an $\mathbb{R}^k$ bundle $NM$ is a map $g:M \to P$ along with an \emph{isomorphism} of bundles $TM \oplus NM \simeq g^*\rho$. For example, the cobordism category appropriate for oriented $n$-dimensional TQFTs is given by taking $P = BSO(n)$ and $\rho$ to be the universal $\mathbb{R}^n$ bundle over $BSO(n)$.

Our goal is to construct $(P,\rho)$ out of $X, G, \xi$ such that a crystalline gauge field on $M$ is the same as a $(P,\rho)$ structure on $M$. Then we will invoke the Baez-Dolan-Lurie cobordism hypothesis (a theorem) to classify TQFTs for such decorated manifolds and compare it to what we have described above.

To do so, consider that, associated to the tangent bundle $TX$, there is a principal $\mathbb{Z}_2$ ($\mathbb{Z}_2 \times B\mathbb{Z}_2$) bundle of orientations (spin structures), of which a section is equivalent to an orientation (spin structure) of $TX$. Let us denote this bundle $Str(TX)$, the structure bundle of $TX$, with the appropriate structure understood for whether we are dealing with bosonic or fermionic systems.

Since $G$ acts smoothly on $X$, the action extends to an action on $TX$, and we can define the $\mathbb{R}^n$ bundle $TX//G \to X//G$, which extends the tangent bundle of $X$ (a fiber of $X//G \to BG$). Likewise we define $Str(TX//G)$ as the bundle of orientations (spin structures).

Now, given a representation $\xi$ of $G$, which represents the action of $G$ on the internal degrees of freedom, we obtain a principal bundle $Str(\xi)$ over $BG$. We can pull this bundle back to $X//G$ using the projection $\pi:X//G \to BG$ to form $\pi^* Str(\xi)$.

Because the structure groups of the two principal bundles $Str(TX//G)$ and $\pi^*Str(\xi)$ are the same and abelian (stable), we can tensor them, to form the $\mathbb{Z}_2$ ($\mathbb{Z}_2 \times B \mathbb{Z}_2$) bundle $Str(TX//G) \otimes \pi^*Str(\xi)$.

Associated to any such bundle is a bundle whose fiber is $BSO(n)$ ($BSpin(n)$). This can be constructed universally, over the classifying space of such bundles, namely $B\mathbb{Z}_2$ ($B\mathbb{Z}_2 \times B^2 \mathbb{Z}_2$). Indeed, these classifying spaces are Postnikov truncations of $BO(n)$, and $O(n)$ acts on both $SO(n)$ and $Spin(n)$. Note however in the case of fermions there are two natural choices of $BSpin(n)$ bundle over $B\mathbb{Z}_2 \times B^2 \mathbb{Z}_2$, depending on whether we take as classifying map $w_2$ or $w_2 + w_1^2$, ie. whether we use $Pin^\pm(n)$. We can decide once and for all to take $Pin^+$ if we agree that the components of the classifying map for $Str(V):Y \to B\mathbb{Z}_2 \times B^2 \mathbb{Z}_2$ are $w_1(V), w_2(V)$, where $V$ is a vector bundle over a space $Y$.

Thus we let $P$ be the $BSO(n)$ ($BSpin(n)$) bundle associated to $Str(TX//G) \otimes \pi^*Str(\xi)$ and we take $\rho$ to be the $\mathbb{R}^n$ bundle which is the universal bundle over all the fiber $BSO(n)$'s ($BSpin(n)$'s). Again this bundle is constructed once and for all over the universal $BSO(n)$ ($BSpin(n)$) bundle over $B\mathbb{Z}_2 \times B^2 \mathbb{Z}_2$, such that the first factor acts by orientation-reversal and the second factor acts trivially. This concludes the construction of the bordism category.

Intuitively, by our construction a $(P,\rho)$ structure on an $n$-manifold $M$ is a map $f:M \to X//G$ as well as a ``discontinuous map" to the fiber $g:M \to BSO(n)$ ($g:M \to BSpin(n)$), whose locus of discontinuity is characterized by the fibration $P$ over $X//G$ with fiber $BSO(n)$ ($BSpin(n)$), which is in turn controlled by the twisting bundle $Str(TX//G) \otimes \pi^*Str(\xi)$. This map is of course fixed by the isomorphism $(f,g)^*\rho \simeq TM$ to be the classifying map of the tangent bundle of $M$, endowing $M$ with the proper twisted tangent structure.

Now we invoke the cobordism hypothesis, which says that, considering the frame bundle $O(\rho)$, which is an $O(n)$-bundle over $P$, an $n$-dimensional TQFT for $(P,\rho)$ manifolds is the same as an $O(n)$-equivariant map from $O(\rho)$ to the $\infty$-groupoid of fully dualizable objects inside some target $n$-category $\mathcal{C}$.

Although we can only be agnostic about the proper choice of target $n$-category $\mathcal{C}$, note that the $\infty$-groupoid of fully dualizable objects inside $\mathcal{C}$ constitutes the space of framed $n$-dimensional TQFTs, also by the cobordism hypothesis. Let us denote this $\Theta_{fr}$.

Note that $O(n)$ acts only on the fibers of $O(\rho)$. Thus, $O(n)$-equivariant maps from $O(\rho)$ to $\Theta_{fr}$ are the same as sections of the hom bundle $Hom_{O(n)}(O(\rho),\Theta_{fr}) \to X//G$, with fiber $Hom_{O(n)}(F,\Theta_{fr})$ where $F$ is the frame bundle of the universal $\mathbb{R}^n$ bundle over $BSO(n)$ ($BSpin(n)$). Again by the cobordism hypothesis, this fiber is isomorphic to the space of bosonic (fermionic) $n$-dimensional TQFTs $\Theta_{SO}$ ($\Theta_{Spin}$). In conclusion we obtain a bundle with fiber $\Theta$ over $X//G$ whose sections are $n$-dimensional TQFTs for manifolds equipped with a crystalline gauge field, with appropriate twists. This should be compared with our construction of the twisted state bundle above, and amounts to a proof of our characterization of crystalline topological liquids via twisted smooth states in conjecture 2 and its proper generalization in Appendix A.

\section{Enlarging the Unit Cell}\label{s:bigunit}

In this section we show that for any space group $G$ with a translation symmetry, there are subgroups $H < G$ with isomorphic point groups but arbitrarily large unit cell (in each dimension), such that if
\[f: BG \to \Theta_d\]
is not homotopic to a constant map, and $i:BH \to BG$ is the map induced by the inclusion $H \hookrightarrow G$, then
\[f \circ i: BH \to \Theta_d\]
is also not homotopic to a constant map. As we discussed in Section \ref{sec:deformations_physical}, this makes our cellular assumption on our local unitary circuits innocuous, since it means that non-trivial $G$ phases remain non-trivial after enlarging the unit cell to $H$.

First of all, let $G_{pt}$ denote the point group of $G$, $T$ the translation subgroup, and $\Lambda \in \mathbb{R}^d$ a lattice for which $G$ is the spacegroup. $G_{pt}$ is finite of order $|G_{pt}|$. Let $m = 1$ mod $|G_{pt}|$ and $\Lambda_m$ be the sublattice of $\Lambda$ where the unit cell is enlarged by $m$ in every dimension. It is straightforward to show that the subgroup $G_m$ of $G$ of symmetries of $\Lambda_m$ is isomorphic to $G$ with the same point group. Let $T_m$ denote the translation symmetry in $G_m$.

Furthermore, since $H^{>0}(BG_{pt},A)$ is $|G_{pt}|$-torsion for any coefficient group $A$ \cite{brown1982cohomology} and $H^0(BG_{pt},A) = A$, the inclusion $G_m \hookrightarrow G$ induces an isomorphism
\[H^{n}(BG,A) \to H^{n}(BG_m,A),\]
for all $n$ if all torsion in $A$ is coprime to $m$. This is because we have a map of Serre spectral sequences with $E_2$ page
\[H^j(BG_{pt},H^k(BT,A)) \to H^j(BG_{pt},H^k(BT_m,A)),\]
where $j+k = n$ and the map is multiplication of the coefficients by $m^k$. Because $m^k = 1$ mod $|G_{pt}|$, this is the identity map on the $E_2$ page except for possibly the $(0,n)$ part, where we get an isomorphism by our assumption on the torsion. Thus the map converges to an isomorphism.

Taking $A = \mathbb{Z}$ and $n = d+2$ thus proves the result for the  group cohomology phases without orientation-reversing symmetries. If we have orientation-reversing elements of $G_{pt}$, then we should include a twist, which is an action of $G_{pt}$ on $\mathbb{Z}$. This yields an action of $G_{pt}$ on the $H^k(BT,\mathbb{Z})$ and $H^k(BT_m,\mathbb{Z})$ for odd $k$. However, even in this twisted setting, $H^{>0}(BG_{pt},A^{tw})$ is still $|G_{pt}|$-torsion, and $H^0(BG_{pt},\mathbb{Z}^{tw})$ is 2-torsion, hence also $|G_{pt}|$-torsion if $G_{pt}$ has an orientation-reversing symmetry. Thus the argument extends to the twisted case with no issue, proving the result for all group cohomology phases.

This argument more generally produces an injection
\[\Omega^n(BG) \to \Omega^n(BG_m)\]
for any generalized cohomology $\Omega$, so long as $m$ is also 1 modulo the product of all torsion in $\Omega^{\le n}(pt)$, using the induced map of Atiyah-Hirzebruch-Serre spectral sequences which on the $E_2$ page is
\[H^j(BG_{pt},\Omega^k(BT)) \to H^j(BG_{pt},\Omega^k(BT_m)),\]
For spin cobordism this means it is good enough to take $m = 1$ mod $|G_{pt}| 2^{d+1}$. It is also enough to produce an isomorphism for twisted cohomology groups.

We can even extend the argument to the most general case of just some target space $\Theta_d$ as above \emph{with bounded homotopy groups}, $\pi_{\ge l} \Theta_d = 0$ for some $l$ (this can be slightly weakened to $\pi_{\ge l} \Theta_d$ is torsion-free). In particular, this is the case if we take $\Theta_d$ to be the space of $(d+1)$-dimensional TQFTs. Let us restrict our attention to a single component of $\Theta_{d,C} \subset \Theta_d$ since $BG$ and $BG_m$ are connected. Let us assume $f \circ i$ is homotopic to a constant map and argue that it follows that $f$ is also homotopic to a constant map, given the further restriction that $m$ is coprime to all the torsion in $\pi_{\le l} \Theta_{d,C}$.

We use obstruction theory to capture the (based) homotopy class of $f$ and $f \circ i$. The first piece of data is a map
\[f_1:\pi_1 BG = G \to \pi_1 \Theta_{d,C}\]
(the unbased homotopy class begins with a conjugacy class of such maps and all further invariants should be taking up to conjugation to capture the unbased homotopy class). So long as $m$ is coprime to the torsion in $\pi_1 \Theta_{d,C}$, then if $f_1$ is nontrivial, so is the induced map
\[f_1 \circ i_1: \pi_1 BG_m = G_m \to \pi_1 \Theta_{d,C}.\]
Thus, by assumption, $f_1$ must be trivial.

The next piece of data is a class
\[f_2 \in H^2(BG, \pi_2 \Theta_{d,C}).\]
As before, if this is nontrivial, then so is the class of the induced map:
\[(f \circ i)_2 = i^* f_2 \in H^2(BG_m, \pi_2 \Theta_{d,C}).\]
Thus, by assumption, and our lemma above, $f_2$ must also be trivial.

We continue this way up to $l$, showing that all obstruction theoretic invariants of $f$ vanish. It follows that $f$ is homotopic to a constant map. This completes the proof of the claimed result, since $m$ may be taken arbitrarily large given the residue constraints.

We note that it is straightforward to extend our results to the case where there is also an internal symmetry $G_{int}$, so long as $m$ is also coprime to all of the torsion in $H^{<d+1}(BG_{int},\mathbb{Z})$. Furthermore, if there is a nontrivial extension
\[G_{int} \to G_{total} \to G_{space},\]
for instance in the case of magnetic translations, as long as $m$ is 1 modulo the order of the extension class, we will still obtain an injective restriction map. For instance if we have a 1/2 magnetic flux per unit cell, then enlarging the unit cell by an odd factor will not change the magnetic translation group. We can then obtain a map of spectral sequences as above to prove the injection.

Furthermore, as it is just a matter of relativizing all the above arguments, the result also straightforwardly extends to the twisted case, where we are studying sections of a $\Theta_d$-fiber bundle over $BG$. All of the cohomology groups in the obstruction theory argument become twisted cohomology groups, but the argument is the same.

\bibliography{../references}

\end{document}